\documentclass[useAMS,fleqn]{mn2e}
\usepackage[]{natbib}
\usepackage[dvips]{graphicx}
\usepackage{amssymb,amsmath}
\usepackage{multirow}
\usepackage{bm}
\usepackage{times}
\usepackage{relsize}
\usepackage{textcomp} 

\def\vth{{\bm \vartheta}}
\def\Rc{\mathcal{R}}
\def\NiT{N_{\rm i}^{\rm T}}
\def\Nio{N_{\rm i}^{\rm o}}

\def\fii{f_{\rm ii}}

\def\fij{f_{\rm ij}}

\def\fik{f_{\rm ik}}
\def\wijo{\omega_{\rm ij}^{\rm o}}
\def\wij{\omega_{\rm ij}}

\def\Dij{(D_{\rm{i}}D_{\rm{j}})_\theta}
\def\RR{(RR)_\theta}
\def\DiR{(D_{\rm{i}}R)_\theta}
\def\DjR{(D_{\rm{j}}R)_\theta}
\def\Nr{N_{\rm R}}
\def\Ni{N_{\rm i}}
\def\Nj{N_{\rm j}}
\def\fow{f_{\rm 12}}
\def\fwo{f_{\rm 21}}
\def\wowo{\omega_{\rm 12}^{\rm o}}
\def\wooo{\omega_{\rm 11}^{\rm o}}
\def\wwwo{\omega_{\rm 22}^{\rm o}}
\def\Noo{N_{\rm 1}^{\rm o}}
\def\Nwo{N_{\rm 2}^{\rm o}}

\begin{document}
\title[CFHTLenS: Quantifying accurate $n(z)$]{CFHTLenS tomographic weak lensing: Quantifying accurate redshift distributions}
\author[Benjamin et al.]
{\parbox{\textwidth}{Jonathan Benjamin$^{1}$\thanks{jonben@phas.ubc.ca}, Ludovic Van Waerbeke$^{1}$, Catherine Heymans$^{2}$, Martin Kilbinger$^{3,4,5,6}$, Thomas Erben$^{7}$, Hendrik Hildebrandt$^{7,1}$, Henk Hoekstra$^{8,9}$, Thomas D. Kitching$^{2}$, Yannick Mellier$^{4,10}$, Lance Miller$^{11}$, Barnaby Rowe$^{12,13}$, Tim Schrabback$^{7,8,14}$, Fergus Simpson$^{2}$, Jean Coupon$^{15}$, Liping Fu$^{16}$, Joachim Harnois-D{\'e}raps$^{17,18}$, Michael J. Hudson$^{19,20}$, Konrad Kuijken$^{8}$, Elisabetta Semboloni$^{8}$, Sanaz Vafaei$^{1}$, Malin Velander$^{8,11}$}\vspace{0.6cm}\\
\parbox{\textwidth}{$^{1}$University of British Columbia, 6224 Agricultural Road, Vancouver, V6T 1Z1, B.C., Canada. \\
$^2$Scottish Universities Physics Alliance, Institute for Astronomy, University of Edinburgh, Royal Observatory, Blackford Hill, Edinburgh, EH9 3HJ, UK. \\
$^3$CEA Saclay, Service d'Astrophysique (SAp), Orme des Merisiers, B\^at 709, F-91191 Gif-sur-Yvette, France \\
$^4$Institut d'Astrophysique de Paris, CNRS, UMR 7095, 98 bis Boulevard Arago, F-75014 Paris, France \\
$^5$Excellence Cluster Universe, Boltzmannstr. 2, D-85748 Garching, Germany \\
$^6$Universit\"ats-Sternwarte, Ludwig-Maximillians-Universit\"at M\"unchen, Scheinerstr.~1, 81679 M\"unchen, Germany \\
$^7$Argelander Institute for Astronomy, University of Bonn, Auf dem H{\"u}gel 71, 53121 Bonn, Germany \\
$^8$Leiden Observatory, Leiden University, Niels Bohrweg 2, 2333 CA Leiden, The Netherlands \\
$^9$Department of Physics and Astronomy, University of Victoria, Victoria, BC V8P 5C2, Canada\\
$^{10}$Institut d'Astrophysique de Paris, Universit{\'e} Pierre et Marie Curie - Paris 6, 98 bis Boulevard Arago, F-75014 Paris, France \\
$^{11}$Department of Physics, Oxford University, Keble Road, Oxford OX1 3RH, UK \\
$^{12}$Department of Physics and Astronomy, University College London, Gower Street, London WC1E 6BT, U.K. \\
$^{13}$California Institute of Technology, 1200 E California Boulevard, Pasadena CA 91125, USA \\
$^{14}$Kavli Institute for Particle Astrophysics and Cosmology, Stanford University, 382 Via Pueblo Mall, Stanford, CA 94305-4060, USA \\
$^{15}$Institute of Astronomy and Astrophysics, Academia Sinica, P.O. Box 23-141, Taipei 10617, Taiwan \\
$^{16}$Key Lab for Astrophysics, Shanghai Normal University, 100 Guilin Road, 200234, Shanghai, China \\
$^{17}$Canadian Institute for Theoretical Astrophysics, University of Toronto, M5S 3H8, Ontario, Canada \\ 
$^{18}$Department of Physics, University of Toronto, M5S 1A7, Ontario, Canada \\
$^{19}$Department of Physics and Astronomy, University of Waterloo, Waterloo, ON, N2L 3G1, Canada\\
$^{20}$Perimeter Institute for Theoretical Physics, 31 Caroline Street N, Waterloo, ON, N2L 1Y5, Canada}}

\maketitle

\begin{abstract}
The Canada-France-Hawaii Telescope Lensing Survey (CFHTLenS) comprises deep multi-colour ($u^{*}g'r'i'z'$) photometry spanning 154 square degrees, with accurate photometric redshifts and shape measurements. We demonstrate that the redshift probability distribution function summed over galaxies provides an accurate representation of the galaxy redshift distribution accounting for random and catastrophic errors for galaxies with best fitting photometric redshifts $z_{\rm p}<1.3$. 

We present cosmological constraints using tomographic weak gravitational lensing by large-scale structure. We use two broad redshift bins $0.5<z_{\rm p}\leq0.85$ and $0.85<z_{\rm p}\leq1.3$ free of intrinsic alignment contamination, and measure the shear correlation function on angular scales in the range ${\sim 1}-40$ arcmin. We show that the problematic redshift scaling of the shear signal, found in previous CFHTLS data analyses, does not afflict the CFHTLenS data. For a flat $\Lambda {\rm CDM}$ model and a fixed matter density $\Omega_{\rm m}=0.27$, we find the normalisation of the matter power spectrum $\sigma_{\rm 8} = 0.771\pm0.041$. When combined with cosmic microwave background data (WMAP7), baryon acoustic oscillation data (BOSS), and a prior on the Hubble constant from the HST distance ladder, we find that CFHTLenS improves the precision of the fully marginalised parameter estimates by an average factor of $1.5-2$. Combining our results with the above cosmological probes, we find $\Omega_{\rm m} = 0.2762\pm0.0074$ and $\sigma_8 = 0.802\pm0.013$.
\end{abstract}


\section{Introduction}
Weak gravitational lensing by large-scale structure provides valuable cosmological information that can be obtained by analysing the apparent shapes of distant galaxies that have been coherently distorted by foreground mass \citep{2001PhR...340..291B}. Since weak lensing is sensitive to the distance-redshift relation and the time-dependent growth of structure, it is a particularly useful tool for constraining models of dark energy \citep{2006astro.ph..9591A, 2006ewg3.rept.....P, 2009arXiv0901.0721A}. To measure the contribution of dark energy over time, the lensing signal must be measured at several redshifts, this is known as weak lensing tomography \citep[see for example,][]{Hu1999,Huterer2002}. Several observations of weak lensing tomography have been completed \citep{2005MNRAS.363..723B,2006A&A...452...51S,2007ApJS..172..239M}. Most recently a study of the COSMOS survey by \citet{Schrabback2010} found evidence for the accelerated expansion of the Universe from weak lensing tomography.

Redshift information is vital to weak lensing interpretation since the distortion of light bundles is a geometric effect and the growth of structure is redshift-dependent. Weak lensing data sets necessitate the use of photometric redshifts due to the large number of galaxies they contain. Spectroscopic redshifts typically exist for a small and relatively-bright fraction of galaxies, providing a training set for photometric redshifts at brighter magnitudes. Several approaches for determining the redshift distribution of galaxies have been used in past weak lensing studies. Many early studies \citep[see for example,][]
{2002A&A...393..369V, Jarvis2003, 2003ApJ...597...98H, 2003MNRAS.344..673B, 2005A&A...429...75V, 2006ApJ...647..116H, Benjamin2007, Fu2008}, lacking multi-band photometry, relied on external photometric redshift samples such as the Hubble Deep Field North and South, and the CFHTLS-Deep fields. Due to the small area of these fields, sampling variance was an important, but often neglected, source of error in these studies, as presented by \citet{2006APh....26...91V}. 

Current and planned weak lensing surveys have multi-band photometry enabling photometric redshift estimates for all galaxies. Methods for measuring photometric redshifts use various model-fitting techniques with the goal of finding a match between the observed photometry and template galaxy spectra which are displaced in redshift and convolved with the optical response of the filter set, telescope, and camera. Depending on the set of photometric filters, degeneracies can exist between different template spectra at different redshifts. We refer to large errors in the best-fitting parameters due to mismatches under these degeneracies as catastrophic errors. The effect of catastrophic errors on weak lensing parameter constraints has been investigated in several studies, for example \citet{Ma2006,BH2010} and \citet{Hearin2010}. Using a detailed Fisher matrix analysis, \citet{Hearin2010} show the importance of properly characterising catastrophic errors to dark energy parameter constraints using weak lensing tomography. The implication of neglecting these errors is not well known, although \citet{Hearin2010} argue that there are many factors governing the final impact on dark energy parameters and each survey needs to be carefully considered to make any definitive statement. It is clear that catastrophic errors will become increasingly important in the next generation of weak lensing cosmic shear surveys. 

In this paper we present a tomographic weak lensing analysis of the Canada-France-Hawaii Telescope Lensing Survey\footnote{http://www.cfhtlens.org} (CFHTLenS), with redshifts measured in \citet{Hendrick2012} using the Bayesian photometric redshift code \citep[{\sc BPZ},][]{2000ApJ...536..571B}. The {\sc BPZ} analysis of the CFHTLenS photometry uses a set of 6 recalibrated spectral energy distribution galaxy templates from \citet{Capak2004} and a magnitude dependent prior on the redshift distribution \citep[see][for further details]{Hendrick2012}. If the galaxy template set and prior used are an accurate and complete representation of the true galaxy population at all redshifts, then the probability distribution function (PDF) calculated using BPZ determines the true error distribution. The redshift distribution of a galaxy sample can then be calculated from the sum of the PDFs to determine an accurate redshift distribution that includes the effects of both statistical and catastrophic errors. This is in contrast to the standard method of using a histogram of photometric redshifts taken from the maximum of the posterior. We test the accuracy of the summed PDFs with overlapping spectroscopic redshifts at bright magnitudes and with resampled COSMOS-30 redshifts \citep{Ilbert2009} at faint magnitudes. In both cases we also assess the level of contamination between redshift bins using an angular cross-correlation technique \citep{2010MNRAS.408.1168B}.

Previous CFHT Legacy Survey (CFHTLS) results were found to be biased, underestimating the shear at high redshifts, thus requiring the addition of a nuisance parameter when model fitting \citep{MK09}. Furthermore, field selection discriminated based on cosmology dependent criterion possibly resulting in confirmation bias. The CFHTLenS catalogues we use in this paper have been thoroughly tested for systematic errors. These tests are cosmology insensitive and were completed without any cosmological analysis of the data \citep{Heymans2012}. One of the primary goals of this paper is to demonstrate that the redshift scaling of the shear is consistent with expectations. We limit our cosmic shear analysis to two broad redshift bins in order to obtain parameter constraints that do not depend on the modelling of intrinsic alignment \citep{Heavens2000, Croft2000, HS04}. A study of cosmological constraints from CFHTLenS with several redshift bins, accounting for intrinsic alignment is presented in \citet{IIGIGG}. \citet{Kilbinger2012} present a thorough investigation of 2D cosmic shear, including a comparison of all popular second order shear statistics. \citet{Simpson2012} use the tomographic shear signal presented in this paper to constrain deviations from General Relativity on cosmological scales.

CFHTLenS has an effective area of 154 square degrees with deep photometry in five broad bands $u^{*}g'r'i'z'$ and a $5\sigma$ point source limiting magnitude in the $i'$-band of ${i'_{\rm AB}{\sim} 25.5}$. These data were obtained as part of the CFHTLS, which completed observations in early 2009. \citet{Heymans2012} present an overview of the CFHTLenS analysis pipeline summarizing the weak lensing data processing with \textsc{THELI} \citep{Erben12}, shear measurement with {\em lens}fit \citep{Miller2012} and photometric redshift measurement from PSF-matched photometry \citep{Hendrick2012} using {\sc BPZ}. Each galaxy in the CFHTLenS catalogue has a shear measurement $\epsilon_{\rm obs}$, an inverse variance weight $w$, a PDF giving the poserior probability as a function of redshift, and a photometric redshift estimate from the peak of the PDF $z_{\rm p}$. The shear calibration corrections described in \citet{Miller2012} and \citet{Heymans2012} are applied and we limit our analysis to the 129 of 171 pointings that have been verified as having no significant systematic errors through a series of cosmology-insensitive systematic tests described in \citet{Heymans2012}.  

This paper is organized as follows, in Section~\ref{sec:sumPDF}, we use a series of tests to determine whether the PDFs are sufficiently accurate to determine the redshift distributions for the many different science analyses of the CFHTLenS data set, and then apply our findings to the first tomographic analysis of the CFHTLenS data set in Section~\ref{sec:tomography}. We investigate the effect of non-linear modelling of the mass power spectrum and baryons on our tomographic weak lensing results in Section~\ref{sec:nonlinear}. Section~\ref{sec:conclusion} contains our concluding remarks.

\section{Galaxy redshift distributions determined from the photometric redshift PDF}
\label{sec:sumPDF}
When considering the redshift of an individual galaxy, a best-fitting redshift must be measured from the PDF, typically corresponding to the peak of the PDF. If many galaxies are considered, the sum of their PDFs can be used as an estimate of the redshift distribution instead of the distribution of best-fitting redshifts. We show in this section that, by using information from the entire PDF, we achieve an accurate model of the redshift distribution. The accuracy of the PDFs is not known {\it a priori} since this depends on whether the template spectral energy distributions and prior information are a representative and complete description of the galaxies in the survey.

We compare the summed PDFs against several other methods of measuring the redshift distribution. These methods include a comparison with the overlapping VVDS and DEEP2 spectroscopic redshifts (see Section~\ref{sec:specz}), statistical resampling of the CFHTLenS photometric redshifts using the COSMOS-30 redshifts (see Section~\ref{sec:COSMOS}), and a photometric redshift contamination analysis (see Section~\ref{sec:contamination}).

We divide the data into six redshift bins and measure the redshift distribution of each. We are limited in the total number of bins by the pairwise contamination analysis, which breaks down for larger numbers of bins (see Section~\ref{sec:contamination} for a more detailed discussion). The redshifts are most reliable in the range $0.1 < z_{\rm p} < 1.3$ where comparison to spectroscopic redshifts, for $i'<24.5$, shows the scatter to be $0.03 < \sigma_{\Delta z} < 0.06$, with an outlier rate of less than 10 per cent \citep{Hendrick2012}. Here $\sigma^2_{\Delta z}$ is the variance in the value of $\Delta z$, which is given by
\begin{equation}
\Delta z = \frac{z_{\rm p}-z_{\rm s}}{1+z_{\rm s}},
\end{equation}
\noindent where $z_{\rm p}$ and $z_{\rm s}$ are the photometric and spectroscopic redshifts, respectively. 

The redshift bins are chosen such that each bin is approximately four times wider than the photometric redshift error $0.04(1+z_{\rm p})$. This is done to avoid excessive contamination between adjacent bins. For $z_{\rm p} > 1.3$ there are only a small number of galaxies, making a subdivision of this range difficult. The six bins are as follows:
\begin{description}
\item[Bin 1:]$0.00 < z_1 \leq 0.17$
\item[Bin 2:]$0.17 < z_2 \leq 0.38$
\item[Bin 3:]$0.38 < z_3 \leq 0.62$
\item[Bin 4:]$0.62 < z_4 \leq 0.90$
\item[Bin 5:]$0.90 < z_5 \leq 1.30$
\item[Bin 6:]$1.30 < z_6$
\end{description}

\begin{figure*}
\centering
\includegraphics[scale=1.00,angle=0]{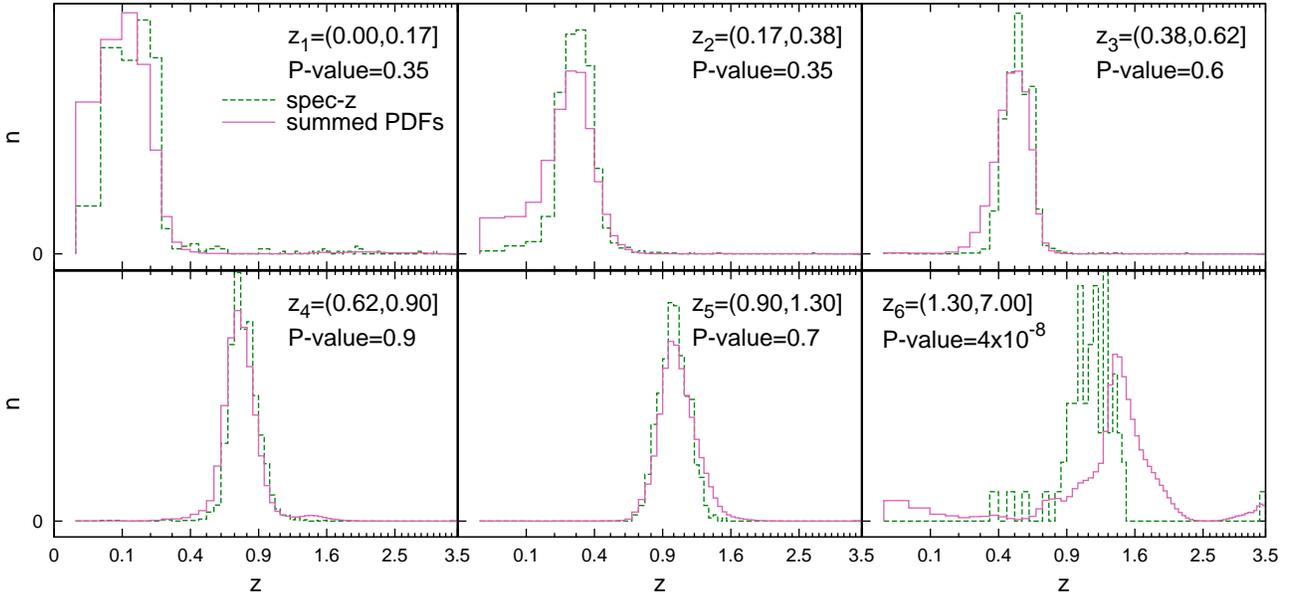}
\caption{\label{fig:contam-1} Comparison of the predicted redshift distributions within each broad redshift bin, labelled $z_{\rm i}$. A magnitude cut of $i' < 23.0$ is used for comparison with spectroscopic redshifts. Solid lines (pink) show the summed PDFs for all galaxies within a given redshift bin. Dashed lines (green) show the spectroscopic redshift distribution. The listed P-values are the result of a two-sample Kolmogorov-Smirnov test of the distributions, we adopt a significance level of $\alpha=0.05$ rejecting the null hypothesis that the two distributions are drawn from the same population for the highest redshift bin.}
\end{figure*}

\subsection{Comparison with spectroscopic redshifts}
\label{sec:specz}
We begin by investigating the redshift distribution given by spectroscopic redshifts. Spectroscopic redshifts from the the VIMOS VLT Deep Survey \citep[VVDS,][]{2005A&A...439..845L} and the DEEP2 galaxy redshift survey \citep{2012arXiv1203.3192N} overlap with CFHTLenS and were used to test the photometric redshifts. For a given photometric redshift bin we can select those galaxies that have spectroscopic redshifts and examine their redshift distribution. The spectroscopic sample is complete for $i' \lesssim 22.0$, dropping to ${{\sim}90}$ per cent completeness for $i' < 23.0$. We adopt the latter cut to ensure that there are a sufficient number of galaxies for our analysis. The catalogues are also cut to exclude objects on masked regions and those that are flagged as stars. Stars are selected with {\tt star\_flag} \citep[see][for more details]{Erben12}. Due to the dithering pattern, which ensures that exposures exist between individual CCD chips, there is a variable number of exposures over a single pointing. This changing photometric depth is difficult to account for when constructing a random catalogue with the same properties, which is necessary for the contamination analysis presented in Section~\ref{sec:contamination}. To avoid this complexity, a final cut is made to select galaxies on areas of the sky that were detected during every exposure, and random objects are placed only in these areas. We do not expect this to bias our results as there is no correlation between the physical properties of a galaxy and which part of the CCD mosaic it was observed on.

A comparison of redshift distributions for $i' < 23.0$ is presented in Figure~\ref{fig:contam-1}. For each redshift bin we show the redshift distribution predicted by the summed PDF (solid line), and the spectroscopic redshift distribution (dashed line). The PDFs of all galaxies within a given redshift bin are summed and the resulting distribution normalised to obtain the solid line. If the summed PDF is a good representation of the true error distribution, then we would expect this distribution to agree with the redshift distribution measured with the spectroscopic redshifts.

We use the Kolmogorov-Smirnov two-sample test (KS test) to determine if the two distributions in Figure~\ref{fig:contam-1} are consistent with being drawn from the same population \citep[details of this test can be found in, for example,][]{Wall}. Before performing the test we adopt as a discriminating criterion a significance level of $\alpha=0.05$. The P-values found from the KS test are presented as labels in Figure~\ref{fig:contam-1}. We find that the distributions for the first five redshift bins are consistent with having been drawn from the same population at a significance level of $\alpha=0.05$. However, we can reject the null hypothesis for the last redshift bin at the same level of significance, indicating that the two distributions are significantly different. This is indicative of the large uncertainties in the photometric redshifts at $z_{\rm p}>1.3$ and confirms the choice of this cut-off made by \citet{Hendrick2012}.

\begin{figure*}
\centering
\includegraphics[scale=1.00,angle=0]{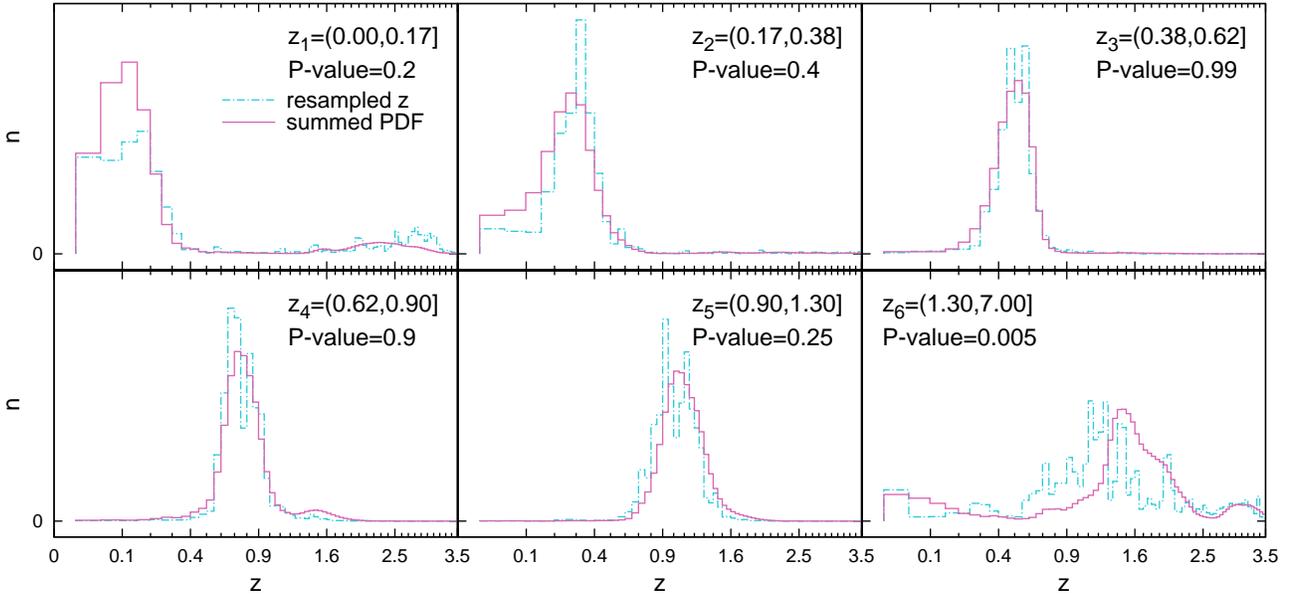}
\caption{\label{fig:contam-2} Comparison of the predicted redshift distributions with a magnitude cut of $i' < 24.7$. Solid lines (pink) show the summed PDFs for all galaxies within a given redshift bin. Dot-dashed histogram (cyan) shows the result of resampling the CFHTLenS redshifts using the constructed conditional probability $P(z_{\rm\mathsmaller{30}}|z_{\rm p})$. The P-values are the result of a KS test, we reject the null hypothesis for the highest redshift bin at $\alpha=0.05$.}
\end{figure*}

\subsection{Comparison with COSMOS photometric redshifts}
\label{sec:COSMOS}
The agreement at bright magnitudes shown in Figure~\ref{fig:contam-1} is encouraging; however, the majority of lensing studies include fainter galaxies, for example a magnitude limit of $i'<24.7$ is adopted for the measurement of CFHTLenS galaxy shapes \citep{Miller2012}. Therefore, we wish to investigate the redshift distribution with this deeper magnitude cut. The spectroscopic redshift sample described in Section~\ref{sec:specz} cannot be used for comparison since the completeness of this sample drops sharply beyond $i' \sim 23$. Instead, we use the COSMOS-30 photometric redshift catalogue \citep{Ilbert2009}, which is accurate to $\sigma_{\Delta z} \simeq 0.012$ due to 30 bands of wavelength coverage from the ultraviolet to the mid-infrared.  Quoting values for the Subaru $i$-band, the COSMOS-30 data are $99.8$ percent complete for $i < 25.5$, and have a $5\sigma$ point source limiting magnitude of $i \sim 26.2$ \citep{Ilbert2009}. A resampling procedure is used to estimate the redshift distribution of deep CFHTLenS galaxies based on the distribution of COSMOS-30 redshifts with CFHTLS overlap.

Although the $1.6$ square degree COSMOS field contains the one square degree CFHTLS-Deep field D2, there are no overlapping CFHTLS-Wide fields. Therefore, it is not possible to directly match CFHTLenS galaxies to objects in the COSMOS-30 catalogue. This issue can be circumvented in a novel way using the photometric catalogue of D2 provided in \citet{Hildebrandt2009}. Using the fact that the photometric systems for the CFHTLS-Wide and Deep data are identical we add random Gaussian noise, scaled to simulate data taken at CFHTLS-Wide depth, to the multi-colour magnitude estimates in the D2 photometric catalogue. Using artificially-degraded catalogues generated in this way, we calculate a Wide-like photometric redshift estimate $z_{\rm p}$ using the maximum of the posterior distribution as described in \citep{Hendrick2012}. This is done for each D2 object in a catalogue matched to the COSMOS-30 catalogue of \citet{Ilbert2009}, employing an association radius of 1.0 arcsec.  The COSMOS-30 redshifts in the matched catalogue are labelled $z_{\rm\mathsmaller{30}}$.

This matched catalogue of $z_{\rm\mathsmaller{30}}$ and noise-degraded Wide-like $z_{\rm p}$ estimates can then be used to acquire information about the joint probability distribution of COSMOS-30 and CFHTLenS redshifts. We generate 100 realisations of the artificially degraded Wide-like catalogues, running the Bayesian photometric redshift estimation of \citet{Hendrick2012} for each realisation. Using this ensemble of $(z_{\rm\mathsmaller{30}}, z_{\rm p})$ pairs, we construct a two-dimensional histogram of galaxy number counts in square bins of width $0.0025$ in redshift for both $z_{\rm\mathsmaller{30}}$ and $z_{\rm p}$.  This histogram is then used as an empirical estimate of the conditional probability density function $P(z_{\rm\mathsmaller{30}} | z_{\rm p})$ and allows us to estimate the corresponding cumulative probability distribution function $P(< z_{\rm\mathsmaller{30}} | z_{\rm p})$ for each $z_{\rm p}$ bin. Then, using inversion sampling from $P(<z_{\rm\mathsmaller{30}} | z_{\rm p})$ with a uniform pseudo-random number generator, samples of redshifts distributed according to $P(z_{\rm\mathsmaller{30}} | z_{\rm p})$ can be drawn.

With the assumption that $P(z_{\rm true} | z_{\rm p}) = P(z_{\rm\mathsmaller{30}} | z_{\rm p})$, the contamination in tomographic redshift bins can be estimated by resampling CFHTLenS redshifts according to $P(z_{\rm\mathsmaller{30}} |z_{\rm p})$. The resulting redshift distributions predicted from this method are given as dot-dash lines in Figure~\ref{fig:contam-2}, and the summed PDFs are presented as solid lines. Note that the fine structure seen in the resampled redshifts is due to structures in the COSMOS field and does not represent real structures in the distribution of CFHTLenS galaxies. The small size of COSMOS means that it is limited by sample variance and individual clusters are able to leave an imprint in the resampled galaxies. This only affects the fine details of the resampled redshifts leading to a breakdown of the assumption $P(z_{\rm true} | z_{\rm p}) = P(z_{\rm\mathsmaller{30}} | z_{\rm p})$ for small redshift intervals.

We again adopt the null hypothesis that the two distributions are drawn from the same population. Using a KS test we find that the null hypothesis can be rejected at a significance level of $\alpha=0.05$ for the $z_{\rm p}>1.3$ redshift bin, but not for any of the other bins. The P-values found from the KS test are presented as labels in Figure~\ref{fig:contam-2}. Our results again confirm that the CFHTLenS photometric redshifts of the $z_{\rm p}>1.3$ galaxies are unreliable. However, we find no evidence that the galaxies at $z_{\rm p}<0.1$, which is the lower limit for the high-confidence redshift range \citep{Hendrick2012}, are unreliable. This is likely because our lowest redshift bin extends to $z_{\rm p}=0.17$ and is therefore dominated by galaxies with well-measured photometric redshifts.

\subsection{Redshift contamination from angular correlation functions}
\label{sec:contamination}
In order to further test the accuracy of the photometric redshift PDFs, we measure the redshift contamination using an angular cross-correlation technique \citep{2010MNRAS.408.1168B}. This method has few assumptions and is sensitive to any contamination between redshift bins. Since it only relies on the angular correlation function of the galaxies it is independent of the other methods used and serves as a critical test.

\subsubsection{Overview of method}
\label{sec:theory-contam}
Galaxies cluster in over-dense regions, leading to an excess in the number of pairs found at a separation $\theta$ when compared to a random distribution of points. The two-point angular correlation function $\omega(\theta)$ quantifies this excess probability of finding pairs. A common estimator \citep{1993ApJ...412...64L} is
\begin{equation}
\wij=\frac{\Dij}{\RR}\frac{\Nr\Nr}{\Ni\Nj} - \frac{\DiR}{\RR}\frac{\Nr}{\Ni}-\frac{\DjR}{\RR}\frac{\Nr}{\Nj} + 1, \label{eq:angcorr_LS}
\end{equation}
\noindent where $\Dij$ is the number of pairs separated by a distance $\theta$ between data sets i and j, $\RR$ is the number of pairs separated by a distance $\theta$ for a random set of points, $\DiR$ is the number of pairs separated by a distance $\theta$ between data set i and a random set of points, $\Nr$ is the number of points in the random sample, and $\Ni$ ($\Nj$) is the number of points in data sample i (j). The auto-correlation is described by the case i$=$j, and the cross-correlation by the case i$\neq$j. Our analysis would hold for any estimator of the angular correlation function.

In the absence of magnification, galaxies in non-overlapping redshift bins should not be clustered with one another. Therefore, clustering between these bins should be consistent with a random distribution of points, resulting in $\wij=0$. Adjacent redshift bins will have a small positive $\wij$ owing to galaxy clustering at their shared edge, which becomes more pronounced for narrow redshift bins. If any non-zero angular cross-correlation is detected between the photometric redshift bins, they must share galaxies with similar redshifts. 

As shown in \citet{2010MNRAS.408.1168B}, this simple realisation can be exploited to estimate contamination between photometric redshift bins. The reader is referred to that work for the full details of the method. Here we present only a few key equations and concepts before we apply the method to the CFHTLenS data.

The contamination fraction, $\fij$ is defined as the number of galaxies contaminating bin j from bin i as a fraction of the total number of galaxies $\NiT$ which have a spectroscopic redshift that lies within redshift bin i. If there is no overlap or contamination between redshift bins, $\fij=0$ when $i \neq j$, and $\NiT = \Nio$, where $\Nio$ is the total number of galaxies which have a photometric redshift that lies within redshift bin i.   In the standard case of overlapping photometric redshift bins, the contamination fraction relates the observed number of galaxies in each photometric redshift bin $\Nio$ to the true underlying number of galaxies $\NiT$ as follows;  
\begin{equation}
\begin{pmatrix}
N_{\rm 1}^{\rm o} \\
N_{\rm 2}^{\rm o} \\
\hdotsfor{1} \\
N_{\rm m}^{\rm o}
\end{pmatrix}=
\begin{pmatrix}
f_{\rm 11} & f_{\rm 21} & \dots & f_{\rm m1} \\
f_{\rm 12} & f_{\rm 22} & \dots & f_{\rm m2} \\
\hdotsfor{4} \\
f_{\rm 1m} & f_{\rm 2m} & \dots & f_{\rm m m} \\
\end{pmatrix}
\begin{pmatrix}
N_{\rm 1}^{\rm T} \\
N_{\rm 2}^{\rm T} \\
\hdotsfor{1} \\
N_{\rm m}^{\rm T}
\end{pmatrix},\label{eq:Niomatrix}
\end{equation}
\noindent where m is the number of redshift bins and $\fii=1-\sum^{\rm m}_{\rm k\neq i}\fik$.  We determine the contamination fractions $\fij$ from measurements of $\wijo$, the observed two-point correlation function between photometric redshift bins i an j. For two redshift bins it can be shown that,
\begin{equation}
 \wowo = \frac{\wooo\left(\frac{\Noo}{\Nwo}\right)\fow(1-\fwo) + \wwwo\left(\frac{\Nwo}{\Noo}\right)\fwo(1-\fow) }{(1-\fow)(1-\fwo) + \fow\fwo}. \label{eq:wijo_2bin}
\end{equation}
When considering more than two redshift bins we measure the contamination fractions for each pair of bins in turn. This pairwise approximation assumes that higher-order contamination can be safely ignored, that is, the angular cross-correlation is not affected by the mutual contamination of the pair of bins by another redshift bin. As the number of bins increases or if the contamination fractions become large, this assumption is no longer valid and the method breaks down.

Once the contamination fractions $\fij$ have been measured, we can invert\footnote{Since we expect the non-diagonal contamination fractions to be small the matrix should be diagonally dominant and therefore invertible.} the contamination matrix in Equation~\ref{eq:Niomatrix} to determine the true underlying number of galaxies in each redshift range $\NiT$ from the observed number of galaxies in each photometric redshift bin $\Nio$. The true redshift distribution $n^{\rm i}(z_{\rm j})$ for each photometric redshift bin i is then calculated over the full redshift range, sampled at each redshift $z_{\rm j}$ from
\begin{equation}
n^{\rm i}(z_{\rm j}) = \fij \NiT . \label{eq:nz}
\end{equation}

\subsubsection{Contamination analysis}
For each pointing the angular correlation function is measured using the publicly available code \textsc{athena}\footnote{http://www2.iap.fr/users/kilbinge/athena/}. \textsc{athena} employs a tree data structure to increase the speed of pair counting at the cost of accuracy. The level of approximation is parametrized by the opening angle. Larger values indicate larger approximations\footnote{Galaxies are grouped together into nodes in the tree data structure based on angular position. The structure is a hierarchy with the nodes on top containing more galaxies. The opening angle determines when to descend to lower nodes and higher spacial resolution.} with an opening angle of zero representing no approximation. Tests of \textsc{athena} against a more simplistic and robust algorithm are used to determine that with an opening angle of 0.03 we are making at most a one per cent error on the angular correlation function. This value of opening angle is used when measuring the angular correlation function.

We measure the angular correlation function for in six angular bins spaced logarithmically on the range $0.15 < \theta < 30$ arcmin. Above $30$ arcmin the signal is very small providing little additional information. For each pointing the contamination fractions are estimated via the angular correlation function as outlined in Section~\ref{sec:theory-contam}. The covariance is estimated via a bootstrap technique, with an additional contribution coming from the field-to-field variance for the angular cross-correlations. The details of the maximum likelihood technique and covariance matrix are presented in Appendix~A of \citet{2010MNRAS.408.1168B}. The likelihoods for the contamination fractions from each field are then combined with equal weighting.

The following matrices contain the measured contamination fractions with 68 per cent confidence regions. All values are multiplied by one hundred for ease of viewing. For the bright sample, $i'<23.0$, we find,
\begin{small}
\begin{equation}
\fij=
\begin{pmatrix}
65\pm 5 &  4\pm 1 & <1      & <1      & <1      &  6\pm 6 \\
28\pm 4 & 87\pm 3 &  8\pm 2 & <1      & <1      &  7\pm 7 \\
 1\pm 1 &  7\pm 2 & 85\pm 3 &  9\pm 2 &  1\pm 1 &  7\pm 7 \\
 1\pm 1 & <1      &  6\pm 2 & 85\pm 2 & 38\pm 6 &  5\pm 5 \\
<1      & <1      & <1      &  4\pm 1 & 56\pm 7 & 29\pm12 \\
 3\pm 1 & <1      & <1      & <1      &  3\pm 1 & 18\pm16 \\
\end{pmatrix}.\label{eq:fmatrix-23.0}
\end{equation}
\end{small}
\noindent With a cut of $i'<24.7$ we measure,
\begin{small}
\begin{equation}
\fij=
\begin{pmatrix}
42\pm 8 &  3\pm 2 & <1      &  1\pm 1 & <1      &  9\pm 6  \\
32\pm 4 & 75\pm 3 &  8\pm 2 & <1      & <1      &  1\pm 1  \\
<1      & 18\pm 2 & 79\pm 3 &  7\pm 2 & <1      &  1\pm 1  \\
 4\pm 4 & <1      & 11\pm 2 & 78\pm 3 & 20\pm 3 &  3\pm 3  \\
<1      & <1      & <1      &  9\pm 2 & 73\pm 4 & 42\pm 5  \\
18\pm 4 &  2\pm 1 & <1      &  4\pm 1 &  5\pm 3 & 36\pm 6  \\
\end{pmatrix}.\label{eq:fmatrix-24.7}
\end{equation}
\end{small}
\noindent Many of the contamination fractions are one per cent deviations from zero, which is expected since we have this level of uncertainty in our estimation of the angular correlation functions. Note that the i$^{\rm th}$ column contains the location of all bin i galaxies. Due to the pairwise treatment of redshift bins columns do not sum to exactly 100 per cent. These matrices are extremely well conditioned with condition numbers of 8.72 and 6.35 for the $i'<23.0$ and $i'<24.7$ cases respectively, indicating that matrix inversion is numerically stable and does not contribute a significant uncertainty to the solution of equation~\ref{eq:Niomatrix}.

With the contamination fractions measured, the true number of galaxies in each redshift bin can be calculated from equation \ref{eq:Niomatrix}. The redshift distribution is then found from equation \ref{eq:nz}. This is done with a Monte Carlo procedure for finding global solutions to the contamination matrix presented in \citet{2010MNRAS.408.1168B}. We can now compare our contamination results with those found from the spectroscopic redshifts and the COSMOS-30 photometric redshifts. However, since our contamination results exist in only six redshift bins we must also sum the distributions shown in Figures~\ref{fig:contam-1}~and~\ref{fig:contam-2} within these six redshift bins.

\begin{figure*}
\centering
\includegraphics[scale=1.00,angle=0]{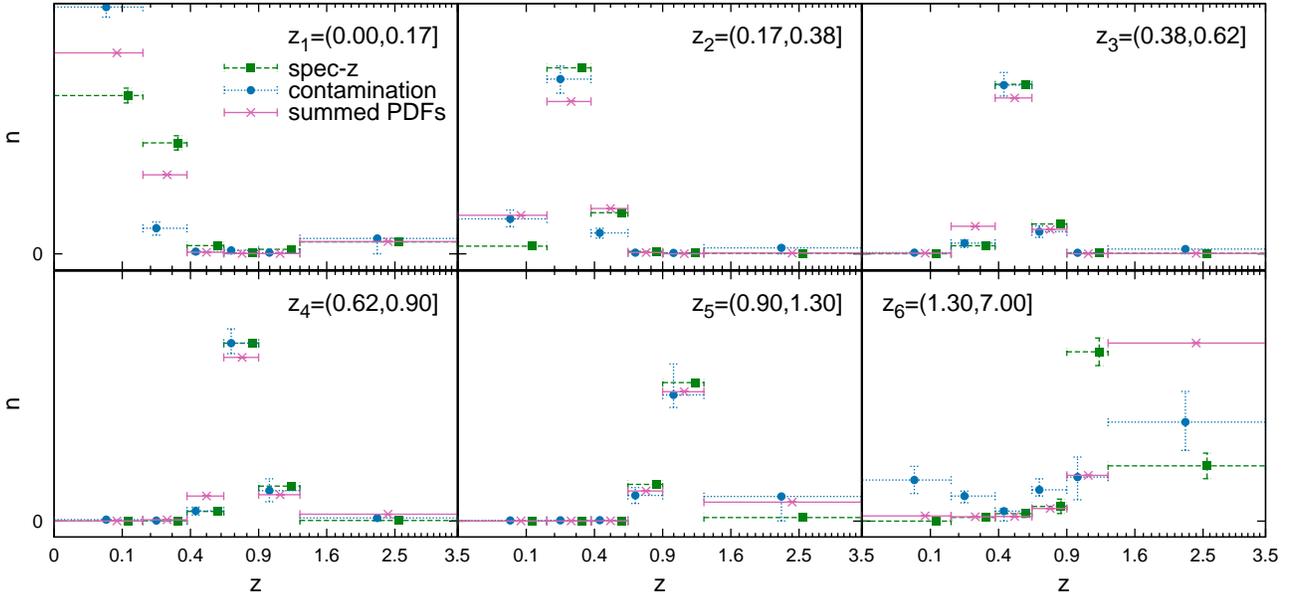}
\caption{\label{fig:contam-23.0} Comparison of the predicted true redshift distribution within each broad redshift bin, labelled $z_{\rm i}$. A magnitude cut of $i' < 23.0$ is used for comparison with spectroscopic redshifts. All horizontal error bars denote the width of the redshift bin and points are offset horizontally for clarity. Crosses with solid lines (pink) denote the summed PDFs when integrated within a given broad redshift bin, the error is calculated as the standard deviation from 1000 bootstrap samples. Filled circles with dotted lines (blue) show the result from our contamination analysis with 68 per cent confidence region. Filled squares with dashed lines (green) show the spectroscopic redshift data integrated within each broad redshift bin. The error is the standard deviation of 1000 bootstrap samples.}
\end{figure*}

\begin{figure*}
\centering
\includegraphics[scale=1.00,angle=0]{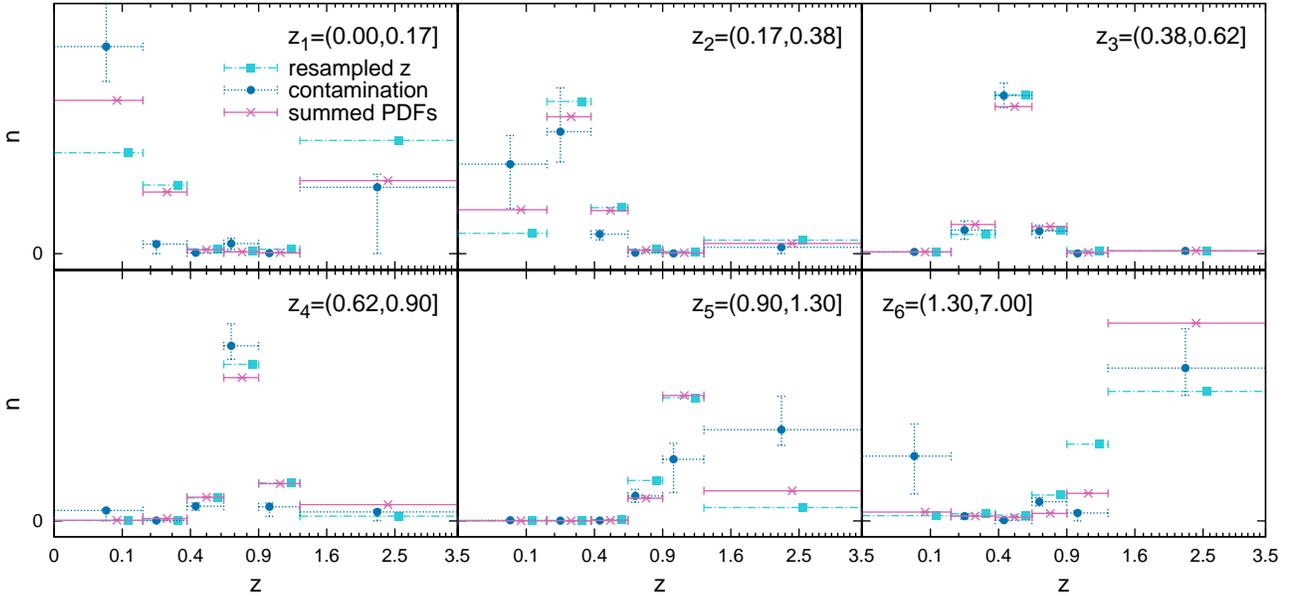}
\caption{\label{fig:contam-24.7} Same as Figure~\ref{fig:contam-23.0} except for the following differences. A magnitude cut of $i' < 24.7$ is used. Filled squares with dot-dashed line (cyan) show the resampled COSMOS-30 data integrated within each broad redshift bin. The error is given as the standard deviation of the 100 low-resolution reconstructions (see Section~\ref{sec:COSMOS}).}
\end{figure*}

For $i'<23.0$ we present our contamination results in Figure~\ref{fig:contam-23.0}. For each redshift bin we show the redshift distribution predicted by the PDFs (crosses with solid lines), the contamination analysis (filled circles with dotted lines), and the spectroscopic redshift distribution (filled squares with dashed lines). The horizontal error bars on all points denote the width of the redshift bins. The vertical scale is proportional to the number of galaxies but uses arbitrary units. For the summed PDF, the vertical error bar is calculated as the standard deviation of the summed PDFs for 1000 bootstraps of the galaxies within each bin $z_{\rm i}$. Note that given the large number of galaxies in each bin (${\sim}40,000-600,000$) the statistical error of the summed PDF is very small. The vertical error bars on the contamination results enclose the 68 per cent confidence region which comes from a procedure for finding global solutions to the contamination \citep{2010MNRAS.408.1168B}. The vertical error on the spectroscopic redshift distribution is taken as the standard deviation from 1000 bootstraps of the spectroscopic catalogue. For both cases where bootstraps are used we verified that 1000 bootstraps yields stable error estimates. For each bootstrap, objects are sampled with replacement and the resulting redshift distributions measured, the total number of galaxies sampled is equal to the number in the original catalogues. 

Figure~\ref{fig:contam-23.0} shows the predicted redshift distribution for each of the six redshift bins used. The contamination points for a given sub-plot are contained within the corresponding row of the contamination matrix in equation~\ref{eq:fmatrix-23.0}. For example, the top row shows that the the majority of galaxies from bin 1 remain in bin 1, with $f_{11}=65\pm5$ per cent. Contamination from other bins is less than the per cent level except for the neighbouring bin $f_{21}=4\pm1$ per cent and the highest redshift bin $f_{61}=6\pm6$. Keep in mind that the relative heights of points in the $z_1$ sub-plot do not follow these contamination values because the contamination $\fij$ represents the number of galaxies in bin j from bin i divided by the true number in bin i. However, investigating the matrix in relation to Figure~\ref{fig:contam-23.0} can help in grasping the presented information. We expect that the spectroscopic redshift distribution is the true distribution, assuming that the limited area of the spectroscopic samples does not bias the results, which is a reasonable assumption for our purposes. The contamination model is in poor agreement with the spectroscopic sample in the $z_1$ and $z_6$ sub-plots. For $z_1$ the contamination model underpredicts the contamination from bin 2 to bin 1, underestimating $f_{21}$ as evidenced by the discrepancy between the contamination point and the spectroscopic point in the second bin of the $z_1$ sub-plot. Similarly the contamination is overpredicted for $f_{12}$ which is seen in the first bin of the $z_2$ sub-plot. This represents a fundamental degeneracy in the angular cross-correlation method. Although an angular cross-correlation is detected between these two bins, unless the angular auto-correlations have significantly different slopes, the method cannot distinguish easily between bin 1 galaxies contaminating bin 2 or vice versa. A similar degeneracy explains the discrepancies in the $z_6$ sub-plot, there we see that $f_{56}$ predicted by the angular cross-correlation is too low and $f_{65}$ in the $z_5$ sub-plot is too high. The contamination between these bins is detected but the direction of scatter is misidentified. 

We use a KS test to determine if the distributions in Figure~\ref{fig:contam-23.0} are consistent with being drawn from the same population. Since there are three distributions and the KS test is a two-sample test we apply it to each pair of distributions. Furthermore, due to the small number of bins we must rely on tabulated critical values which exist for very few significance levels, therefore we are not able to list the P-values for each redshift bin. For each pair of distributions we find that we cannot reject the null hypothesis (drawn from the same population) at a significance level of $\alpha=0.05$ for any of the redshift bins. 

We present the results of the contamination analysis for $i'<24.7$ in Figure~\ref{fig:contam-24.7}. The summed PDF and contamination results are presented similarly to Figure~\ref{fig:contam-23.0}. The resampled redshifts using COSMOS-30 are given as the dot-dash line and filled squares. The vertical error on the COSMOS-30 points is taken as the standard deviation of the 100 low-resolution resamplings (see Section~\ref{sec:COSMOS}). If we compare the contamination results to the resampled redshifts the greatest discrepancies are for the $z_1$, $z_5$, and $z_6$ sub-plots. The $f_{12}-f_{21}$ and $f_{56}-f_{65}$ degeneracies noted for the bright sample above appear again in Figure~\ref{fig:contam-24.7}. Additionally the resampled redshifts predict a larger $f_{61}$ and smaller $f_{16}$ than do the contamination results which can be seen in the first and last bins of the $z_1$ and $z_6$ sub-plots respectively. The contamination analysis predicts that a significantly lower number of galaxies belong in bin 5 compared to the other methods, see bin 5 of the $z_5$ sub-plot. However, this is not due to scattering of bin 5 galaxies elsewhere, note that $f_{55}=73\pm4$ per cent, instead the contamination analysis simply predicts fewer galaxies occupying this bin. To determine if these differences are statistically significant we use a KS test. We again have three distributions and apply the test between each pair. For each pair of distributions we find that we cannot reject the null hypothesis (drawn from the same population) at a significance level of $\alpha=0.05$ for any of the redshift bins. 

When using the finely binned spectroscopic and resampled redshifts in Sections~\ref{sec:specz}~and~\ref{sec:COSMOS} we were able to reject the null hypothesis for the high redshift bin $z_{\rm p}>1.3$. The smallest P-value found was in the high redshift bin when comparing the summed PDF with the spectroscopic distribution in Figure~\ref{fig:contam-1}. Performing the same comparison with these distributions when summed within the six redshift bins of the contamination analysis we are not able reject the null hypothesis. The coarse binning required by the contamination analysis has reduced the statistical power of the test. However, the contamination analysis provides a complementary estimation of the redshift distribution, which agrees well with the other estimates and strengthens our confidence in the summed PDF as an accurate measure of the redshift distribution.

We conclude that the summed PDF can be used to estimate the redshift distribution for the high confidence redshift range $0.1 < z_{\rm p} < 1.3$ determined by \citet{Hendrick2012}. The comparison with the resampled COSMO-30 redshifts and the contamination analysis for $i'<24.7$ suggest that an accurate estimate of the redshift distribution, including statistical and catastrophic errors, can be obtained from the sum of the PDFs. This result suggests that the model galaxy spectra and priors used in \citet{Hendrick2012} are a fair and sufficiently complete representation for the population of galaxies studied here.

\section{Weak lensing tomography}
\label{sec:tomography}
In this paper we present an analysis of the CFHTLenS tomographic weak lensing signal using two broad redshift bins and compare our results with a 2D analysis over the same redshift range.  Setting our analysis in a flat $\Lambda$CDM cosmology framework, the initial aim is to use the consistent results we find between successive tomographic bins as a demonstration that the CFHTLenS catalogues are not subject to the redshift-dependent systematic biases that were uncovered in an earlier analysis of CFHTLS data \citep{MK09}.  This cosmology-dependent demonstration is the last in an extensive series of tests, which investigate the robustness and accuracy of the CFHTLenS catalogues.  We stress, however, that this analysis was performed after the conclusion of a series of cosmology-insensitive tests presented in \citet{Heymans2012} and the photometric redshift accuracy analysis presented in Section~\ref{sec:sumPDF}.  Most importantly, no feedback loop existed between this cosmology-dependent test and the systematics and image simulation tests that determined the calibration corrections and the subset of reliable data that we use from the survey.

We choose to use two broad mid-to-high redshift bins for our tomographic analysis in order to reduce the potential contamination to the signal from intrinsic galaxy alignments \citep[see, for example,][and references therein]{Heavens2000,IIGIGG}. We estimate the expected contamination of the measured weak lensing signal using the linear tidal field intrinsic alignment model of \citet{HS04}, and following \citet{BK07} by fixing its amplitude to the observational constraints obtained by \citet{BTHD02}.  By limiting the redshift bins to photometric redshifts $0.5 < z_{\rm p} \leq 0.85$ and $0.85 < z_{\rm p} \leq 1.3$, we estimate that any contamination from intrinsic alignments is expected to be no more than a few per cent for each redshift bin combination.  We therefore ignore any contributions from intrinsic alignments in this analysis as they are expected to be small in comparison to our statistical errors.  Note that a low level of contamination would not be expected if we instead used the 6 narrow redshift bins that were analysed in the redshift contamination analysis in Section~\ref{sec:sumPDF}.   We present a fine 6-bin tomographic analysis of the data in \citet{IIGIGG} where the impact of intrinsic galaxy alignments is mitigated via the simultaneous fit of a cosmological model and an intrinsic alignment model. The findings of \citet{IIGIGG} support the approach taken in this paper to neglect the contribution of intrinsic alignments for our choice of redshift bins.

The 2D lensing analysis presented here is restricted to the same redshift range used in our tomographic analysis $0.5 < z_{\rm p} \leq 1.3$. We measure the shear correlation function on angular scales from ${\sim}1$ to ${\sim}40$ arcmin. The upper limit is set by our ability to measure the covariance matrix from simulations, see Section~\ref{sec:covar}.

\subsection{Overview of tomographic weak lensing theory}
\label{sec:wltheory}
The complex weak lensing shear $\gamma=\gamma_1 + {\rm i}\gamma_2$, which is directly analogous to the complex galaxy ellipticity, can be decomposed into two components: the tangential shear $\gamma_{\rm t}$ and the cross component $\gamma_{\rm x}$. These are defined relative to the separation vector for each pair of galaxies, with $\gamma_{\rm t}$ describing elongation and compression of the ellipticity along the separation vector and $\gamma_{\rm x}$ describing elongation and compression along a direction rotated $45^{\circ}$ from the separation vector. The following shear-shear correlation functions can then be computed:
\begin{equation}
\xi_\pm^{\rm k,l}(\theta)=\frac{\Sigma_{\rm i,j}\left[ \gamma_{\rm t,i}^{\rm k}(\vth_{\rm i})\gamma_{\rm t,j}^{\rm l}(\vth_{\rm j}) \pm \gamma_{\rm x,i}^{\rm k}(\vth_{\rm i}) \gamma_{\rm x,j}^{\rm l}(\vth_{\rm j}) \right]w_{\rm i}w_{\rm j}\Delta_{\rm ij}}{\Sigma_{\rm i,j} w_{\rm i}w_{\rm j}\Delta_{\rm ij}},
\label{xipm}
\end{equation}
\noindent where galaxy pairs labelled {\rm i,j} are separated by angular distance $\vartheta = |\vth_{\rm i} - \vth_{\rm j}|$. If $\vartheta$ falls in the angular bin given by $\theta$ then $\Delta_{\rm ij}$=1, otherwise $\Delta_{\rm ij}$=0. The labels {\rm k,l} identify redshift bins. The summation is performed for all galaxies i in bin k and all galaxies j in bin l. The contribution of each galaxy pair is weighted by its inverse variance weight $w_{\rm i}w_{\rm j}$. This gives greater significance to galaxy pairs with well-measured shapes. 

Shear calibration is performed as described in \citet{Miller2012} and \citet{Heymans2012}. This signal-to-noise (S/N) and size-dependent calibration includes an additive ($c$) and a multiplicative ($m$) correction term as follows;
\begin{equation}
\gamma^{\rm obs}=(1+m)\gamma^{\rm true}+c.
\end{equation}
\noindent An average additive correction of $2\times10^{-3}$ is found for $\gamma_2$. The additive correction for $\gamma_1$ is found to be consistent with zero. The multiplicative correction to $\xi_{\pm}$ is found by calculating the weighted correlation function of $1+m$ \citep{Miller2012},
\begin{equation}
1+{\rm K}^{\rm k,l}(\theta)=\frac{\Sigma_{\rm i,j}(1+m_{\rm i}^{\rm k})(1+m_{\rm j}^{\rm l})w_{\rm i}w_{\rm j}\Delta_{\rm ij}}{\Sigma_{\rm i,j} w_{\rm i}w_{\rm j}\Delta_{\rm ij}}.
\end{equation}
\noindent The shear correlation functions $\xi_{\pm}$ are then corrected by dividing them by $1+{\rm K}$.

The shear-shear correlations can also be expressed as filtered functions of the convergence power spectra
\begin{equation}
\xi_{+/-}^{\rm k,l}(\theta)={\frac{1}{2\pi}} \int_0^\infty~{\rm d}\ell \, \ell \, J_{0/4}(\ell \theta)P_\kappa^{\rm k,l}(\ell),
\label{theogg}
\end{equation}
\noindent where $J_{\rm n}$ is the ${\rm n^{th}}$ order Bessel function of the first kind and $\ell$ is the modulus of the two-dimensional wave vector. These can be related to line-of-sight integrals of the three-dimensional matter power spectrum

\begin{equation}
P_\kappa^{\rm k,l}(\ell)=\frac{9\,H_0^4\,\Omega_{\rm m}^2}{4c^4}\int_0^{\chi_{\rm_h}} {\rm d}\chi {g_{\rm k}(\chi)g_{\rm l}(\chi) \over a^2(\chi)} P_{\delta}\left({\ell \over f_K(\chi)},\chi \right),
\label{pofkappa}
\end{equation}
\noindent where $c$ is the speed of light, $\Omega_{\rm m}$ is the matter energy density, $H_0$ is the Hubble constant, $f_K(\chi)$ is the comoving angular diameter distance out to a distance $\chi$, $\chi_{\rm_h}$ is the comoving horizon distance, $a(\chi)$ is the scale factor, and $P_{\delta}$ is the 3-dimensional mass power spectrum computed from a non-linear estimation of dark matter clustering \citep{Smith2003}. The two terms, $g_{\rm k}(\chi)$, are the geometric lens-efficiency, which depend on the redshift distribution of the sources, $n_{\rm k}(\chi')$,
\begin{equation}
g_{\rm k}(\chi) = \int_\chi^{\chi_{\rm_h}}{\rm d} \chi' n_{\rm k}(\chi') {f_K(\chi'-\chi)\over f_K(\chi')}.
\label{lensgeom}
\end{equation}

Given a cosmological model, matter power spectrum, and redshift distribution of the sources we can model the shear correlation functions. Bayesian model fitting techniques are then used to obtain the posterior probability on the model vector given the observed shear correlation functions. We discuss this further in Section~\ref{sec:cosmology}.

\subsection{The tomographic weak lensing signal}
\label{sec:signal}

\begin{figure}
\centering
\includegraphics[scale=1.3,angle=0]{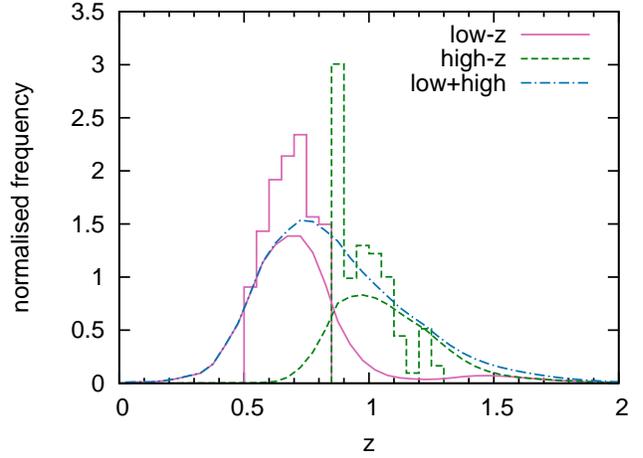}
\caption{\label{fig:nz_2bin} Redshift distributions used in the weak lensing analysis. Low and high redshift bins correspond to $z_{\rm p}=(0.5,0.85]$ and $z_{\rm p}=(0.85,1.3]$ respectively. Smooth curves show the result of summing the photometric redshift probability distribution functions (PDFs) of all galaxies within the respective redshift bin, and the solid and dashed curves are used in the tomographic analysis. The sum of the PDFs over the entire redshift range is given by the dot-dashed line which is used in the 2D lensing analysis. For comparison, the histograms show the redshift distribution obtained from the photometric redshifts.}
\end{figure}

\begin{figure}
\centering
\includegraphics[scale=1.2,angle=0]{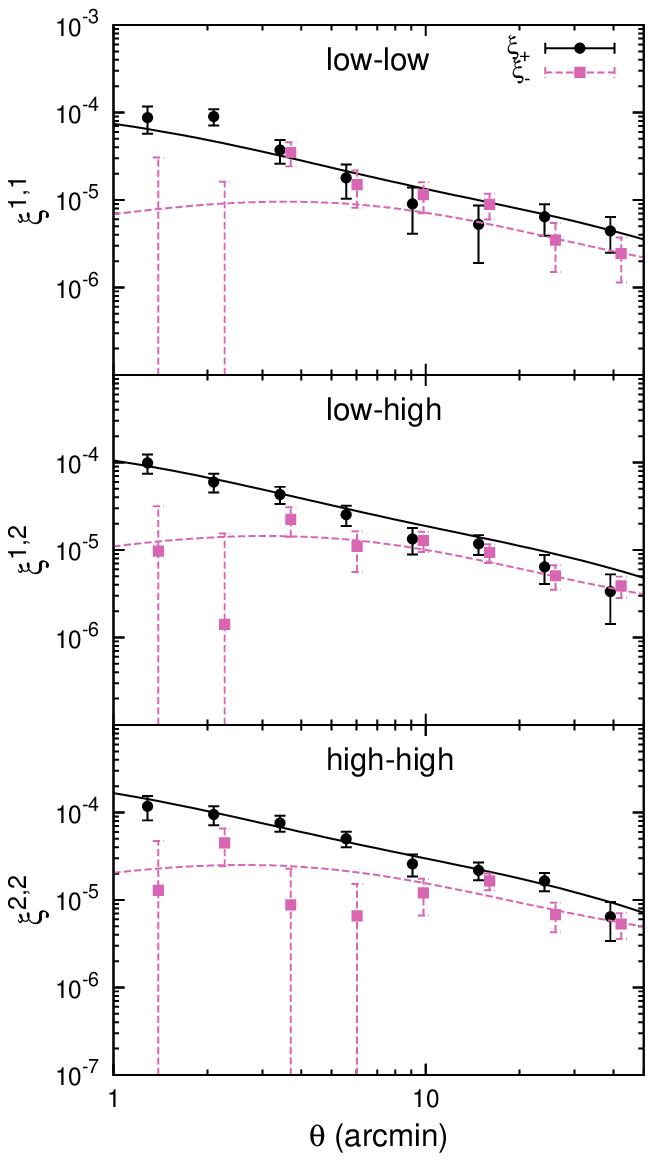}
\caption{\label{fig:signal} The filled circles with  solid lines and filled squares with dashed lines show the measured signal for $\xi_{+}$ and $\xi_{-}$ respectively. Each panel shows the shear correlation functions for a unique pairing of redshift bins. The top, middle and bottom panels correspond to low redshift correlated with low redshift (low-low), low with high redshift (low-high), and high with high redshift (high-high). Error bars are the square-root of the diagonal of the covariance matrix measured from mock catalogues (see Section~\ref{sec:covar}). Theoretical predictions for a fiducial \protect\citep[WMAP7,][]{WMAP7} cosmology are presented as lines; these are not the best-fitting models. There are two negative data points for $\xi_{-}$ in the top panel, their values are $-2.3\times10^{-6}$ and $-4.9\times10^{-6}$ for scales 1.34 and 2.18 arcminutes respectively.}
\end{figure}

Based on the results presented in Section~\ref{sec:sumPDF}, the redshift distribution in each bin is taken to be the sum of the PDFs determined from the photometric redshift analysis of \citet{Hendrick2012}. We refer to the maximum posterior photometric redshift estimate as the `photometric redshift'. The histogram of photometric redshifts and the sum of the PDFs for each redshift bin are presented in Figure~\ref{fig:nz_2bin}. Note that the summed PDFs extend to lower and higher redshifts then the photometric redshifts do, broadening the range below $z=0.5$ and above $z=1.3$. The summed PDFs for the two redshift bins also overlap considerably with one another. The average redshift from the summed PDFs is 0.7 for the low redshift bin and 1.05 for the high redshift bin. For the photometric redshifts we find 0.69 and 1.03 for the low and high redshift bins respectively. The average redshift for both bins taken together is found to be 0.87 from the summed PDFs and 0.84 from the photometric redshifts.

\begin{table*}
\caption{Details of the model dependent cosmological parameters for each of the considered cosmologies. Parameter ranges denote hard priors. A flat distribution is used throughout the range. The bottom three parameters are constrained by WMAP7, and are required in order to deduce $\sigma_{\rm 8}$.}
\label{tab:cosmoParams}
\begin{center}
\begin{tabular}{lllll}
\hline
\hline
Parameter & flat $\Lambda {\rm CDM}$ & curved $\Lambda {\rm CDM}$ & description \\
\hline
$\Omega_{\rm m}$ & $[0,1.0]$ & $[0,1.2]$ & Energy density of matter (baryons + dark matter). \\
$\sigma_8$ & $[0.2,1.5]$ & $[0.2,1.5]$ & Normalisation of the matter power spectrum.  \\
$\rm h$ & $[0.4,1.2]$ & $[0.4,1.2]$ & The dimensionless Hubble parameter ${\rm h}=\frac{{\rm H}_{0}}{100 {\rm km \, s^{-1} Mpc^{-1}}}$. \\
$\Omega_{\rm b}$ & $[0,0.1]$ & $[0,0.1]$ & Energy density of baryons. \\
$n_{\rm s}$ & $[0.7,1.3]$ & $[0.7,1.3]$ & Slope of the primordial matter power spectrum. \\
$\Omega_{\Lambda}$ & $1-\Omega_{\rm m}$ & $[0,2]$ & Energy density of dark energy. \\
$w_{0}$ & $-1$ & $-1$ & Constant term in the dark energy equation of state, $w(a)=w_{0}$. \\
\hline
$\tau$ & $[0.04,0.20]$ & $[0.04,0.20]$ & Reionisation optical depth. \\
$\Delta^2_{\Rc}$ & $[1.8,3.5]$ & $[1.8,3.5]$ & Amplitude of curvature perturbations, units of $10^{-9}$ times the amplitude of density fluctuations. \\
${\rm A_{SZ}}$ & $[0.0,2.0]$ & $[0.0,2.0]$ & Sunyaev-Zel{\textquotesingle}dovich template amplitude. \\
\hline
\hline
\end{tabular}
\end{center}
\end{table*}

We use \textsc{athena} with an opening angle of $0.02$ to measure the shear-shear correlation function. We have tested that the difference to the shear-shear correlation function when using an opening angle of $0.02$ compared to a brute force calculation is negligible, approximately $8$ per cent of the size of the errors. The signal is first measured on each of the four wide mosaics: W1,W2,W3, and W4, applying the shear calibration described in Section~\ref{sec:wltheory}. The correlation functions are then combined by calculating the weighted average. The weight for a given angular bin and wide mosaic is the sum of the inverse variance weight terms for each pair of galaxies. We present $\xi_{+}$ and $\xi_{-}$ for each redshift bin combination in Figure~\ref{fig:signal}. The error bars correspond to the diagonal elements from the covariance matrix, discussed in more detail in Section~\ref{sec:covar}. The lines are the theoretical prediction for a fiducial cosmological model using WMAP7 best-fitting results \citep{WMAP7}, hence the following parameter vector is used: ($\Omega_{\rm m}=0.271$, $\sigma_8=0.78$, $h=0.704$, $\Omega_{\rm b}=0.0455$, $n_{\rm s}=0.967$, $\Omega_{\Lambda}=0.729$, $w_{0}=-1$). Descriptions of each parameter can be found in Table~\ref{tab:cosmoParams}. To compute the theoretical models we employ the halo-model of \citet{Smith2003} to estimate the non-linear matter power spectrum and the analytical approximation of \citet{1998ApJ...496..605E} to estimate the transfer function.

Emphasizing that no cosmology-dependent systematic tests were used to vet the catalogues \citep{Heymans2012}, Figure~\ref{fig:signal} demonstrates the robustness of the CFHTLenS catalogues. The tomographic shear signal shows no evidence of a redshift-dependent bias as was seen in earlier CFHTLS data analyses \citep{MK09}. We discuss further tests of the redshift scaling of the shear in Section~{\ref{sec:zscale}}.

\subsection{Cosmology}
\label{sec:cosmology}
From the signal measured in Section~\ref{sec:signal}, cosmological parameters are estimated using \textsc{cosmoPMC}. \textsc{cosmoPMC} is a freely available\footnote{http://cosmopmc.info} Population Monte Carlo (PMC) code, which uses adaptive importance sampling to explore the posterior likelihood \citep{arXiv1101.0950K}. \textsc{cosmoPMC} documentation can be found in \citet{arXiv1101.0950K}; discussion of Bayesian evidence and examples of its application to various cosmological data sets can be found in \citet{2010MNRAS.405.2381K} and \citet{2009PhRvD..80b3507W}. The PMC method is detailed in \citet{2007arXiv0710.4242C}. The non-linear matter power spectrum is estimated using the halo-model of \citet{Smith2003}. The transfer function is estimated using the analytical approximation of \citet{1998ApJ...496..605E}.

We explore two cosmologies: a flat $\Lambda {\rm CDM}$ universe and a curved $\Lambda {\rm CDM}$ universe. The model-dependent data vector used with \textsc{cosmoPMC} contains the following seven parameters: $(\Omega_{\rm m}, \sigma_8, {\rm h}, \Omega_{\rm b}, n_{\rm s}, \Omega_{\Lambda}, w_0)$. Physical descriptions and priors are presented in Table~\ref{tab:cosmoParams}. For the flat $\Lambda {\rm CDM}$ model we have a five-parameter fit, where we fix $\Omega_{\Lambda} = 1-\Omega_{\rm m}$ and $w_0=-1$. For the curved $\Lambda {\rm CDM}$ model we have six free parameters as $\Omega_{\Lambda}$ is allowed to vary, while $w_0$ remains fixed.
 
Three other cosmological data sets are used to provide complementary constraining power. Constraints from the cosmic microwave background (CMB) are taken from the seven-year results of WMAP \citep[][hereafter referred to as WMAP7]{WMAP7}. Baryon acoustic oscillation (BAO) data is taken from the BOSS experiment \citep[][hereafter referred to as BOSS]{BOSS}. The Hubble constant is constrained with the results from the HST distance ladder \citep[][hereafter referred to as R11]{R11}. Following R11, we use a Gaussian prior of mean value $H_0=0.738$ and standard deviation $\sigma=0.024$. For more details of these data sets see \citet{Kilbinger2012}. With WMAP7 the parameter set is expanded to include $\tau$, ${\rm A_{SZ}}$, and $\Delta^2_{\Rc}$, from which we deduce $\sigma_{\rm 8}$. Prior ranges and brief descriptions are given in Table~\ref{tab:cosmoParams}. For further details see \citet{WMAP7} and references therein. Throughout this section when stating parameter values we quote the 68.3 per cent confidence level as the associated uncertainty with all other parameters marginalised over.

\subsubsection{Covariance Matrix}
\label{sec:covar}
In order to estimate a covariance matrix for our measured shear correlation functions in equation~\ref{xipm}, we analyse mock CFHTLenS surveys constructed from the three-dimensional N-body numerical lensing simulations of \citet{Clone2012}.   The $1024^3$ particle simulations have a box size of $147.0 \, {\rm h}^{-1} {\rm Mpc}$ or $231.1 \, {\rm h}^{-1} {\rm Mpc}$, depending on the redshift of the simulation, and assume a flat $\Lambda$CDM cosmology parametrized by the best-fitting constraints from \citet{WMAP5}.  There are a total of 184 fully independent lines of sight spanning 12.84 square degrees with a resolution of 0.2 arcmin sampled at 26 redshift slices between $0<z<3$. The two-point shear statistics match the theoretical predictions of the input cosmology from $0.5 < \theta < 40$ arcmin scales at all redshifts \citep{Clone2012}, this sets the upper angular limit for our tomographic analysis.  See \citet{IIGIGG} for a detailed discussion of covariance matrix estimation from the N-body simulations presented in \citet{Clone2012}, including the required \citet{Anderson} correction that we apply to de-bias our estimate of the inverse covariance matrix used in the likelihood analysis that follows.

\subsubsection{Flat $\Lambda {\rm CDM}$}
\label{sec:flatLCDM}
\begin{table}
\caption{Constraints orthogonal to the $\Omega_{\rm m}-\sigma_{\rm 8}$ degeneracy for a flat $\Lambda {\rm CDM}$ cosmology. Results are shown with and without highly non-linear scales which are potentially biased due to non-linear modelling and the effects of baryons (see Section~\ref{sec:nonlinear}). `All scales' refers to scales the correlation functions are measured on: $1 < \theta < 40$ arcmin. We remove scales corresponding to $\xi_- < 10$ arcmin in the case labelled `removed: $\xi_- < 10$ arcmin'.}
\label{tab:alpha}
\begin{center}
\begin{tabular}{lcc}
\hline
Data & $\sigma_{\rm 8}\left(\frac{\Omega_{\rm m}}{0.27}\right)^{\alpha}$ & $\alpha$ \\
\hline
\hline
tomography: & & \\
all scales & $0.771\pm0.040$ & $0.553\pm0.016$ \\ 
removed: $\xi_- < 10$ arcmin & $0.776\pm0.041$ & $0.556\pm0.018$ \\ 
\hline
\hline
2D Lensing: & & \\
all scales & $0.785\pm0.036$ & $0.556\pm0.018$ \\ 
removed: $\xi_- < 10$ arcmin & $0.780\pm0.043$ & $0.611\pm0.015$ \\ 
\hline
\end{tabular}
\end{center}
\end{table}

\begin{figure}
\centering
\includegraphics[scale=0.65,angle=0]{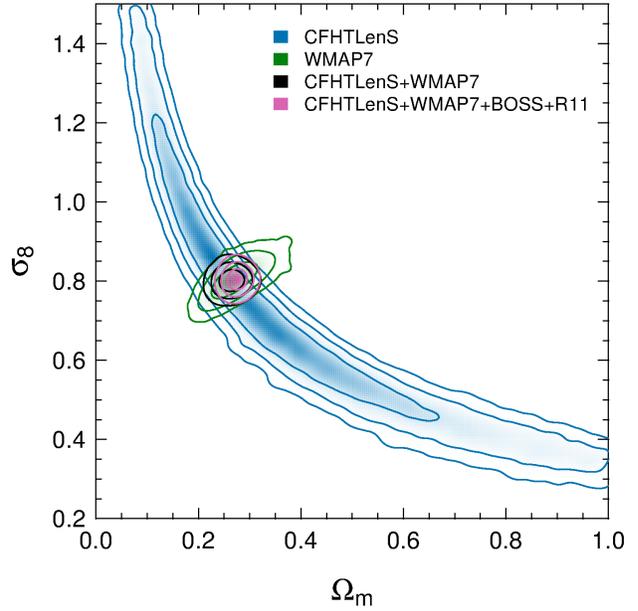}
\caption{\label{fig:om_sig}Marginalised parameter constraints (68.3, 95.5, and 99.7 per cent confidence levels) in the $\Omega_{\rm m}-\sigma_{\rm 8}$ plane for a flat $\Lambda$CDM model. Results are shown for CFHTLenS (blue), WMAP7 (green), CFHTLenS combined with WMAP7 (black), and CFHTLenS combined with WMAP7, BOSS and R11 (pink).}
\end{figure}

\begin{table}
\caption{Parameter constraints with 68.3 per cent confidence limits. The following parameters are deduced for CFHTLenS: $\Omega_{\rm K}$ and $q_0$. When combining data sets the deduced parameters are: $\sigma_{\rm 8}$, $\Omega_{\Lambda}$, and $q_0$. The label CFHTLenS+Others refers to the combination of CFHTLenS, WMAP7, BOSS, and R11.}
\label{tab:cosmo}
\hspace{-0.44cm}
\resizebox{8.8cm}{!} {
\begin{tabular}{@{}l@{}@{}c@{}@{}c@{}@{}l@{}}
\hline
Parameter & flat $\Lambda {\rm CDM}$ & curved $\Lambda {\rm CDM}$ & Data \\
\hline
\hline
\multirow{3}{*}{$\Omega_{\rm m}$}
 & $0.27\pm0.17$ & $0.28\pm0.17$ & CFHTLenS\\ 
 & $0.288\pm0.010$ & $0.285\pm0.014$ & WMAP7+BOSS+R11\\ 
 & $0.2762\pm0.0074$ & $0.2736\pm0.0085$ & CFHTLenS+Others\\ 
\hline
\multirow{3}{*}{$\sigma_{\rm 8}$}
 & $0.67\pm0.23$ & $0.69\pm0.29$ & CFHTLenS\\ 
 & $0.828\pm0.023$ & $0.819\pm0.036$ & WMAP7+BOSS+R11\\
 & $0.802\pm0.013$ & $0.795\pm0.013$ & CFHTLenS+Others\\
\hline
\multirow{3}{*}{$\Omega_{\Lambda}$}
 & $1-\Omega_{\rm m}$ & $0.38\pm0.36$ & CFHTLenS\\ 
 & $1-\Omega_{\rm m}$ & $0.717\pm0.019$ & WMAP7+BOSS+R11\\
 & $1-\Omega_{\rm m}$ & $0.7312\pm0.0094$ & CFHTLenS+Others\\
\hline
\multirow{3}{*}{$\Omega_{\rm K}$}
 & $0$ & $0.19\pm0.43$ & CFHTLenS\\ 
 & $0$ & $-0.0020\pm0.0061$ & WMAP7+BOSS+R11\\
 & $0$ & $-0.0042\pm0.0040$ & CFHTLenS+Others\\
\hline
\multirow{3}{*}{$h$}
 & $0.84\pm0.25$ & $0.81\pm0.24$ & CFHTLenS\\ 
 & $0.692\pm0.0088$ & $0.694\pm0.012$ & WMAP7+BOSS+R11\\
 & $0.6971\pm0.0081$ & $0.693\pm0.011$ & CFHTLenS+Others\\
\hline
\multirow{3}{*}{$\Omega_{\rm b}$}
 & $0.030\pm0.029$ & $0.031\pm0.030$ & CFHTLenS\\ 
 & $0.0471\pm0.0012$ & $0.0472\pm0.0016$ & WMAP7+BOSS+R11\\
 & $0.04595\pm0.00086$ & $0.0470\pm0.0015$ & CFHTLenS+Others\\
\hline
\multirow{3}{*}{$q_0$}
 & $-0.57\pm0.27$ & $-0.29\pm0.40$ & CFHTLenS\\ 
 & $-0.568\pm0.016$ & $-0.574\pm0.025$ & WMAP7+BOSS+R11\\
 & $-0.585\pm0.011$ & $-0.594\pm0.014$ & CFHTLenS+Others\\
\hline
\multirow{3}{*}{$n_s$}
 & $0.93\pm0.17$ & $0.91\pm0.17$ & CFHTLenS\\ 
 & $0.965\pm0.012$ & $0.969\pm0.014$ & WMAP7+BOSS+R11\\
 & $0.960\pm0.011$ & $0.972\pm0.012$ & CFHTLenS+Others\\
\hline
\multirow{2}{*}{$\tau$}
 & $0.086\pm0.014$ & $0.086\pm0.015$ & WMAP7+BOSS+R11\\
 & $0.081\pm0.013$ & $0.085\pm0.015$ & CFHTLenS+Others\\
\hline
\multirow{2}{*}{$\Delta^2_{\Rc}$}
 & $2.465\pm0.086$ & $2.45\pm0.13$ & WMAP7+BOSS+R11\\
 & $2.429\pm0.081$ & $2.361\pm0.094$ & CFHTLenS+Others\\
\hline
\multirow{2}{*}{$A_{\rm SZ}$}
 & $0.97\pm0.62$ & $1.35\pm0.61$ & WMAP7+BOSS+R11\\
 & $1.33\pm0.60$ & $1.39\pm0.57$ & CFHTLenS+Others\\
\hline
\end{tabular}}
\end{table}

We present marginalised two-dimensional likelihood constraints in the $\Omega_{\rm m}-\sigma_{\rm 8}$ plane in Figure~\ref{fig:om_sig}. The best constraint from weak lensing alone is for a combination of $\Omega_{\rm m}$ and $\sigma_{\rm 8}$, which parametrizes the degeneracy. We find $\sigma_{\rm 8}\left(\frac{\Omega_{\rm m}}{0.27}\right)^{\alpha}=0.771\pm0.040$ with $\alpha=0.553\pm0.016$. 

When combining CFHTLenS with WMAP7, BOSS, and R11 data sets, we find $\Omega_{\rm m}=0.2762\pm0.0074$ and $\sigma_{\rm 8}=0.802\pm0.013$. The precision is ${\sim}20$ times better than for CFHTLenS alone where we find $\Omega_{\rm m}=0.27\pm0.17$ and $\sigma_{\rm 8}=0.67\pm0.23$. Constraints on the full set of parameters are presented in Table~\ref{tab:cosmo}. We show the results for CFHTLenS tomography, CFHTLenS combined with WMAP7, BOSS and R11, and, to assess the contribution of our data set to these constraints, we include results for WMAP7 combined with R11 and BOSS. The most valuable contribution from CFHTLenS is for $\Omega_{\rm m}$, $\sigma_8$, and $\Omega_{\rm b}$, where we improve the precision of the constraints by an average factor of $1.5$.

\begin{figure}
\centering
\includegraphics[scale=0.65,angle=0]{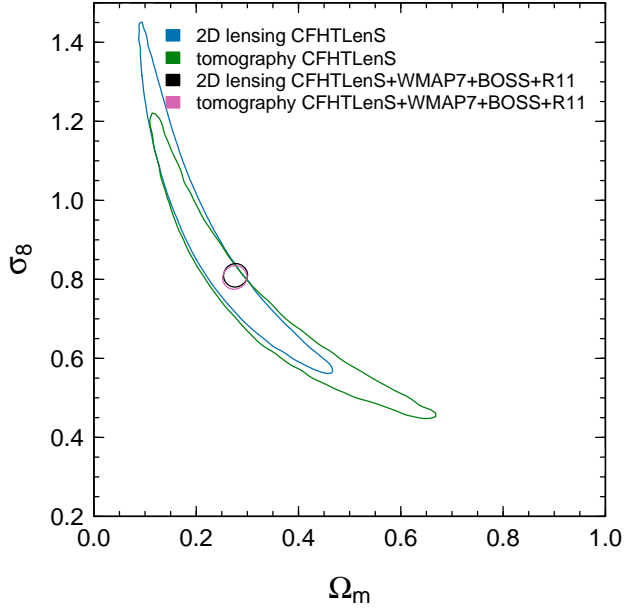}
\caption{\label{fig:1bin_comp}Marginalised parameter constraints (68.3 per cent confidence level) in the $\Omega_{\rm m}-\sigma_{\rm 8}$ plane for a flat $\Lambda$CDM cosmology. We compare the results for 2D lensing (blue) and 2-bin tomography (green). We combine CFHTLenS with WMAP7, BOSS, and R11. Results are shown for 2D lensing (black) and 2-bin tomography (pink).}
\end{figure}

For comparison we perform the analysis with a single redshift bin spanning the range of our 2-bin analysis, $0.5 < z_{\rm p} \leq 1.3$. We refer to this as the 2D lensing case, in contrast to the tomographic case where we split the galaxies into two redshift bins.  Figure~\ref{fig:1bin_comp} shows the marginalised parameter constraints in the $\Omega_{\rm m}-\sigma_{\rm 8}$ plane for both 2D lensing and tomography, and the two cases result in very similar constraints. For 2D lensing we find $\sigma_{\rm 8}\left(\frac{\Omega_{\rm m}}{0.27}\right)^{\alpha}=0.785\pm0.036$ and $\alpha=0.556\pm0.018$, which is in agreement with what we find for tomography (Table~\ref{tab:alpha}). When combining the 2D lesning results from CFHTLenS with WMAP7, BOSS, and R11 data sets, we find $\Omega_{\rm m}=0.2774\pm0.0074$ and $\sigma_{\rm 8}=0.810\pm0.013$, which are nearly identical to those found for tomography (listed above and in Table~\ref{tab:cosmo}). This level of agreement is also found for all other parameters when combining CFHTLenS with the other data sets. We note that all the parameter estimates agree within the 68.3 per cent errors and the size of the error bars from 2D lensing are very similar to those found with tomography when combining CFHTLenS with WMAP7, BOSS, and R11 data sets.

For CFHTLenS alone the parameter estimates for 2D lensing and tomographic lensing agree with each other within their 68.3 per cent uncertainties, however, the parameter constraints do not improve. With two broad overlapping redshift bins of average redshift $0.7$ and $1.05$, there appears to be insufficient additional information to tighten parameter constraints. Previous estimates of the improvement in constraints from weak lensing tomography use non-overlapping redshift bins, Gaussian covariance, and estimate errors using a Fisher matrix analysis \citep[see][]{Simon2004}. For two redshift bins with $z<3$ and divided at $z=0.75$ they find the error on individual parameters from tomography to be 88 per cent those from a 2D analysis. With our overlapping redshift bins and non-Gaussian covariance it is not surprising that this marginal improvement is significantly degraded. Additionally, our 2D lensing result is for redshifts $0.5<z_{\rm p}<1.3$, this removes low redshift galaxies with small signal-to-noise improving constraints and weakening the gains from dividing the redshift range. We therefore expect a modest improvement at best. We find that our tomographic errors are on average 116 per cent of our 2D lensing errors. The fact that we find larger errors for tomography is surprising and warrants further discussion. 

To test our covariance matrices we perform the analysis again, replacing the measured shear correlation function with that predicted from our model using a WMAP7 cosmology. In this case we find that the tomographic errors are 98.2 per cent of the 2D errors. Therefore the increase in the tomographic errors compared to the 2D errors is not an inherent product of the covariance matrices used. This result confirms that for the case of overlapping redshift bins and non-Gaussian covariance the expected improvement from tomography is marginal, at best.

As discussed in detail in section~\ref{sec:nonlinear} we have also analysed the data after removing small scales which could be affected by errors in the non-linear modelling of the matter power spectrum and baryonic effects. When removing these scales ($\xi_-<10$ arcmin) we see an improvement in the errors for tomography which are measured to be 104 per cent those of 2D lensing with the same scales removed. The remaining discrepancy could be due to several factors. The \citet{Smith2003} non-linear prescription could easily be biased at the few per cent level. Residual errors in the redshift of galaxies or other per cent level systematics could be present. In addition we expect some degradation of the tomography errors due to the bias correction of the inverse covariance matrix \citep{Anderson}. The covariance is estimated from a finite number of mock catalogues (see Section~\ref{sec:covar}), since the tomographic covariance contains three times the number of elements as the 2D covariance, measuring it from the same number of mock catalogues results in a noisier measure. \citet{Hartlap} predict an erroneous increase in likelihood area of three per cent given our number of mock catalogues 184 and the size of the data vector for 2D lensing 16 and tomography 48.

Finally we note that our two redshift bins are chosen based on concerns of intrinsic alignment contamination, as such, they are not optimised for constraining cosmology. With more carefully selected redshift bins it may be possible to overcome the issues discussed above and obtain improved cosmological constraints.

\subsubsection{Redshift scaling of the cosmic shear signal}
\label{sec:zscale}
Previous CFHTLS data were found to underestimate the shear signal at high redshift necessitating additional calibration parameters when performing cosmological fits to the data \citep{MK09}. We demonstrate here that the CFHTLenS data have a redshift dependent shear-signal which agrees with expectations from the modelled $\Lambda$CDM cosmology.

The excellent agreement between the 2D and tomographic lensing results (Figure~\ref{fig:1bin_comp}) suggests that the shear signal across our two redshift bins is scaling as expected. This is also observed in the excellent agreement between the measured shear and the shear prediction based on a fiducial WMAP7 cosmology shown in Figure~\ref{fig:signal}.

\begin{figure}
\centering
\includegraphics[scale=0.65,angle=0]{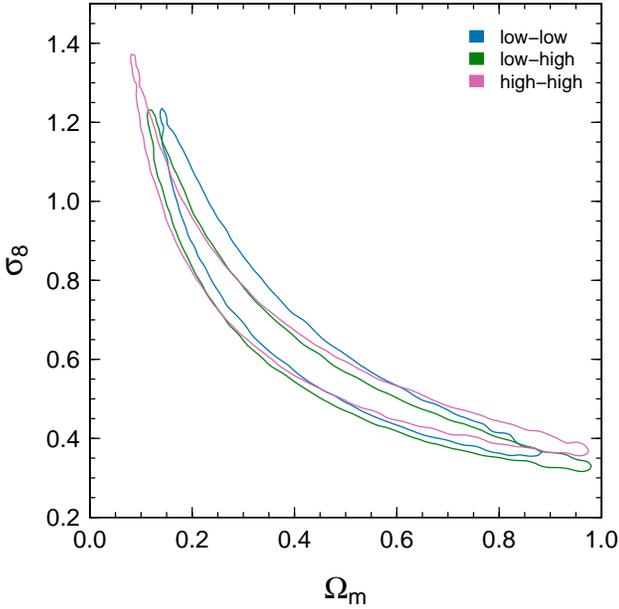}
\caption{\label{fig:zscale}Marginalised parameter constraints (68.3 per cent confidence level) in the $\Omega_{\rm m}-\sigma_{\rm 8}$ plane for a flat $\Lambda$CDM cosmology. The results are shown for each combination of the two redshift bins. The low and high redshift bins correspond to $0.5 < z_{\rm p} \leq 0.85$ and $0.85 < z_{\rm p} \leq 1.3$ respectively. The excellent agreement shows that redshift scaling of the signal is consistent with the modelled $\Lambda$CDM cosmology.}
\end{figure}

The shear correlation function for each pair of tomographic redshift bins is analysed separately, corresponding to the shear correlation functions shown in each panel of Figure~\ref{fig:signal}. In Figure~\ref{fig:zscale} we present marginalised parameter constraints (68.3 per cent confidence level) in the $\Omega_{\rm m}-\sigma_{\rm 8}$ plane for each redshift bin combination. Since each contour is obtained from a sub-sample of the full data-set the degeneracy between the parameters is more pronounced and the area of the contours is larger than when analysing the full data set (Figure~\ref{fig:om_sig}). The agreement between the contours in Figure~\ref{fig:zscale} is a convincing demonstration that the redshift scaling of the shear in the CFHTLenS data is consistent with expectations from the modelled $\Lambda$CDM cosmology.

The power-law fits to the degenerate parameter constraints in Figure~\ref{fig:zscale} for each case are $\sigma_{\rm 8}\left(\frac{\Omega_{\rm m}}{0.27}\right)^{\alpha}=0.820\pm0.067,\,0.753\pm0.053,\,{\rm and}\,0.753\pm0.050$ with $\alpha=0.662\pm0.020,\,0.621\pm0.016,\,{\rm and}\,0.535\pm0.013$ for the low-low, low-high, and high-high redshift bin pairings respectively. Note the expected evolution of $\alpha$ with redshift.

We reiterate that the cosmological model-dependent verification of redshift scaling presented here is completely independent of the calibration of the data, and the rejection of bad fields, that were done with tests which are not sensitive to cosmology \citep{Heymans2012}. 

\subsubsection{Curved $\Lambda {\rm CDM}$}
\begin{figure}
\centering
\vbox{
\includegraphics[scale=0.65,angle=0]{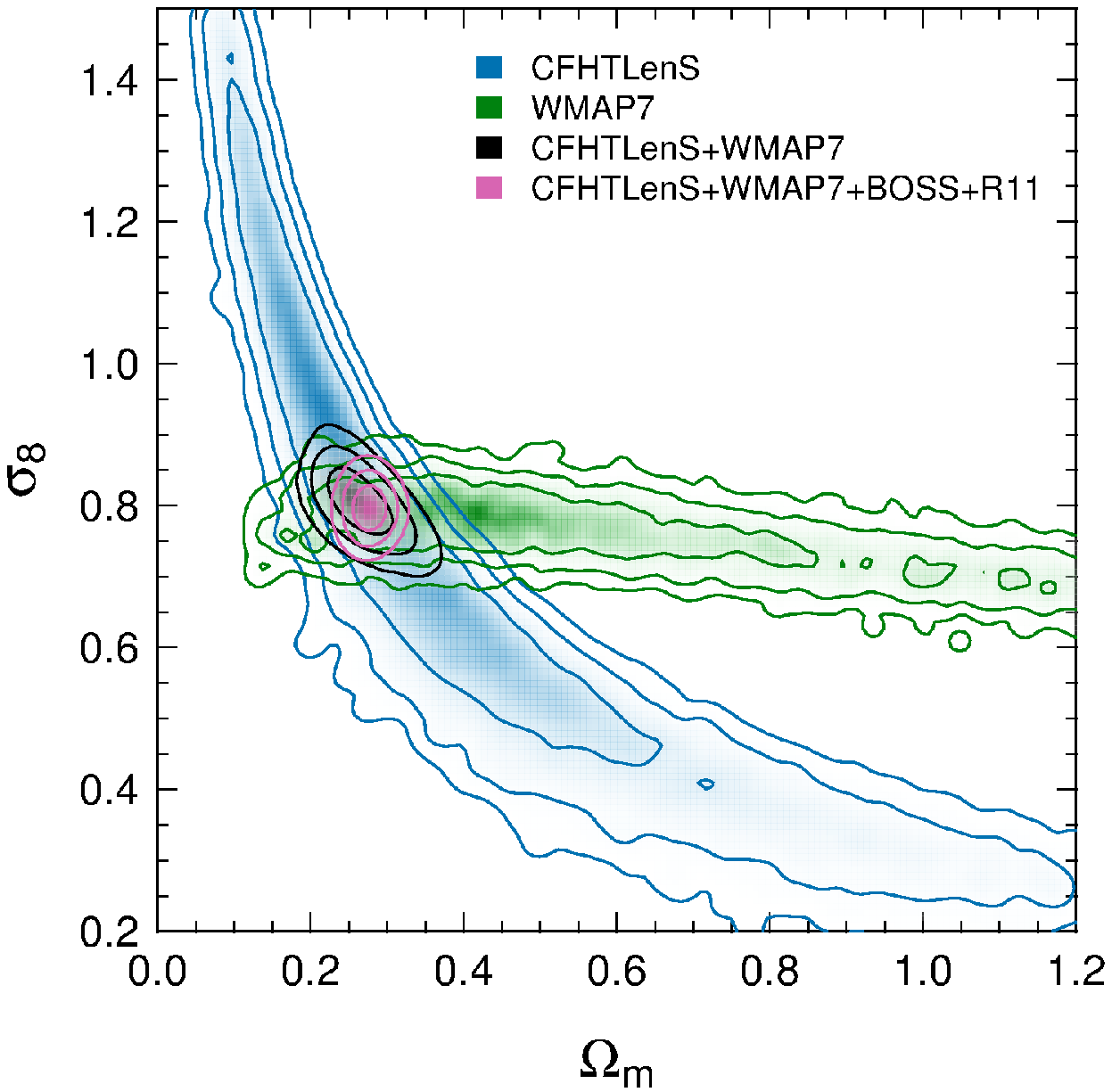}
\includegraphics[scale=0.65,angle=0]{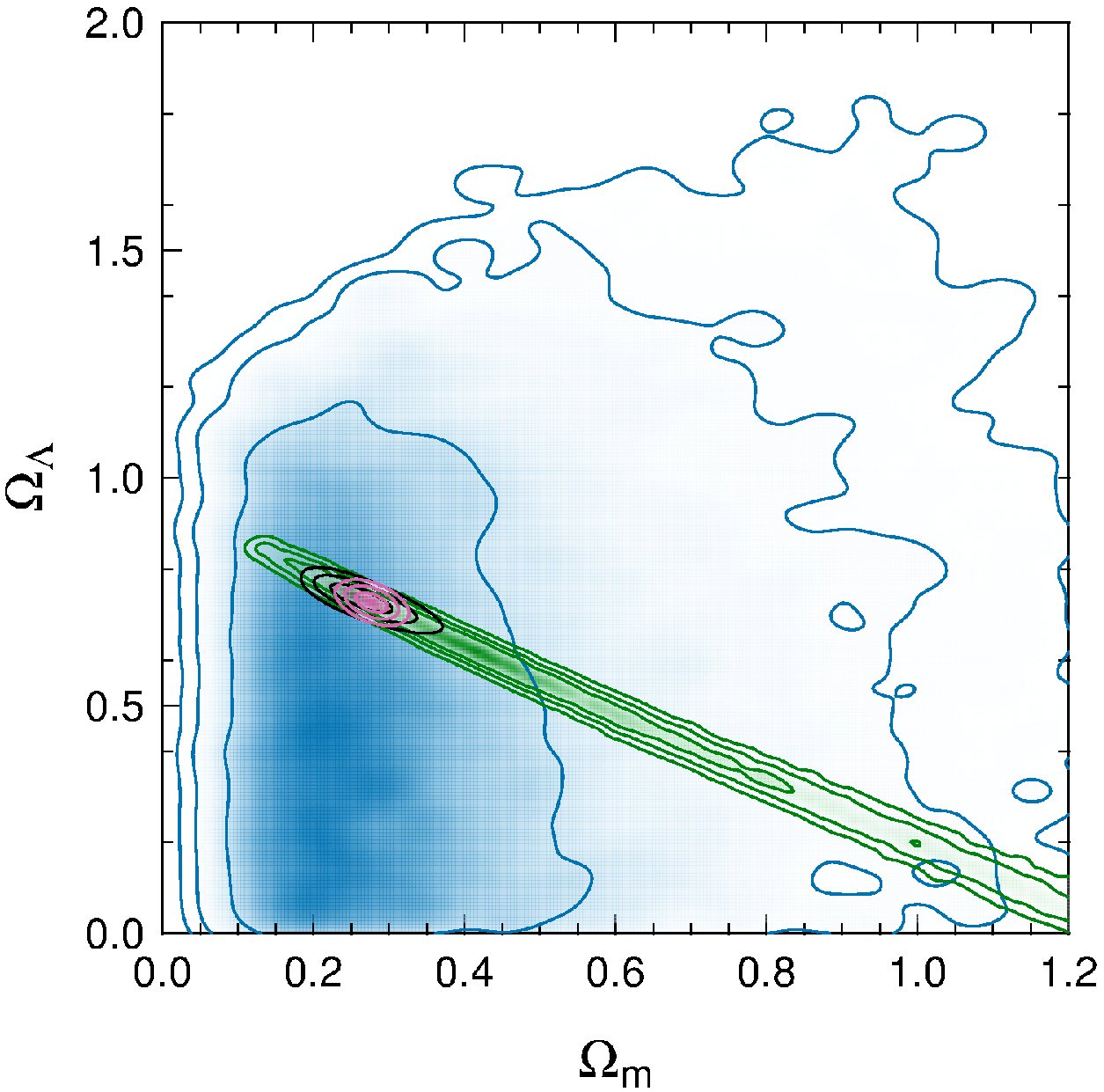}
}
\caption{\label{fig:om_ode} Marginalised parameter constraints (68.3, 95.5, and 99.7 per cent confidence levels) for a curved $\Lambda$CDM cosmology. Results are shown for CFHTLenS (blue), WMAP7 (green), CFHTLenS combined with WMAP7 (black), and CFHTLenS combined with WMAP7, BOSS, and R11 (pink). {\bf Top panel:} Constraints in the $\Omega_{\rm m}-\sigma_{\rm 8}$ parameter space.  {\bf Bottom panel:} Constraints in the $\Omega_{\rm m}-\Omega_{\Lambda}$ parameter space.}
\end{figure}

A curved $\Lambda {\rm CDM}$ cosmology is modelled, for the full details of parameters and priors used see Table~\ref{tab:cosmoParams}. We present constraints in the $\Omega_{\rm m}-\sigma_8$ and $\Omega_{\rm m}-\Omega_{\Lambda}$ plane in Figure~\ref{fig:om_ode}. One-dimensional marginalised results when combining CFHTLenS with WMAP7, BOSS, and R11 are $\Omega_{\rm m}=0.2736\pm0.0085$, $\Omega_{\Lambda}=0.7312\pm0.0094$, $\Omega_{\rm K}=-0.0042\pm0.0040$, and $\sigma_{\rm 8}=0.795\pm0.013$. The constraints on $\Omega_{\rm m}$ and $\sigma_8$ do not change significantly from the flat $\Lambda {\rm CDM}$ case. Parameter constraints for both models are presented in Table~\ref{tab:cosmo}. The addition of CFHTLenS to WMAP7, BOSS, and R11 is most helpful at constraining $\Omega_{\rm m}$, $\sigma_8$, $\Omega_{\rm K}$, and $\Omega_{\Lambda}$. The precision for these parameters improves, on average, by a factor of two.

We again find excellent agreement with the 2D lensing analysis. When combining the 2D lensing of CFHTLenS with WMAP7, BOSS, and R11 data sets, we find $\Omega_{\rm m}=0.2766\pm0.0082$, $\Omega_{\Lambda}=0.7273\pm0.0089$, $\Omega_{\rm K}=-0.0035\pm0.0035$ and $\sigma_{\rm 8}=0.804\pm0.016$. We do not show the complete details of our 2D lensing parameter estimations. However, we note that in all cases, either with CFHTLenS alone or combined with the other cosmological probes, the 2D results agree with the tomographic results within the 68.3 per cent errors and the size of the error bars are similar for both cases. We again find that for CFHTLenS alone the individual parameter uncertainties from the tomographic analysis are 116 per cent those of the 2D lensing analysis.

\subsubsection{Constraining the deceleration parameter}
\begin{figure}
\centering
\includegraphics[scale=0.65,angle=0]{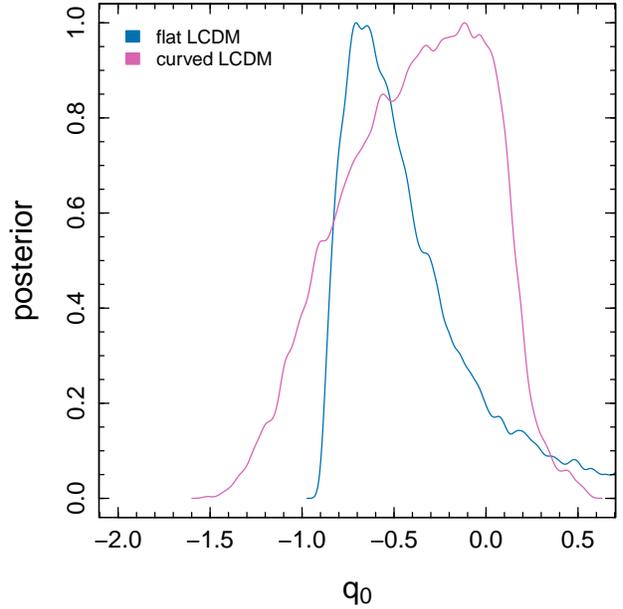}
\caption{\label{fig:qacc_LCDM}Marginalised constraints on the deceleration parameter using the CFHTLenS 2-bin tomographic weak lensing results. An accelerating universe ($q_0 < 0$) is found at the 82 per cent confidence level for a curved $\Lambda$CDM model (pink), and at the 89 per cent confidence level for a flat $\Lambda$CDM model (blue).}
\end{figure}

The deceleration parameter $q_0$ parametrizes the change in the expansion rate of the Universe. We calculate this as a deduced parameter for both the flat and the curved $\Lambda$CDM models. The deceleration parameter depends on the energy density parameters
\begin{eqnarray}
q_0&\equiv&-\frac{\ddot{a}(t_0)a(t_0)}{\dot{a}^2(t_0)}=\frac{\Omega_{\rm m}}{2} - \Omega_{\Lambda}\;\; ({\rm curved}\, \Lambda{\rm CDM}),\; {\rm and}\nonumber \\
&=&\frac{3\Omega_{\rm m}}{2} - 1\;\;  ({\rm flat}\, \Lambda{\rm CDM}),
\end{eqnarray}
\noindent where the scale factor at present time is $a(t_0)$ and derivatives with respect to time are denoted with a dot. For the flat case $q_0$ is simply a transformation of our results for the matter density parameter $\Omega_{\rm m}$. We present marginalised constraints for $q_0$ in Figure~\ref{fig:qacc_LCDM}. The pink line is for the curved case where we find $q_0=-0.29\pm0.40$, and the blue line is for the flat case where we find $q_0=-0.57\pm0.27$. Negative values indicate acceleration of the expansion of the Universe. Summing the posterior for $q_0<0$ tells us the confidence level at which we have measured an accelerating Universe. For the curved and flat models we find that $q_0<0$ at the 82 and 89 per cent confidence level, respectively.

\citet{Schrabback2010} constrain $q_0$ with a six-bin tomographic analysis of the COSMOS-30 data. Besides having more tomographic bins the redshift range probed is also greater, extending to $z=4$. For a curved $\Lambda$CDM cosmology, holding $\Omega_{\rm b}$ and $n_s$ fixed and using a Gaussian prior on the Hubble constant of $h=0.72\pm0.025$, they find $q_0<0$ at 96 per cent confidence. If we do a similar analysis with $\Omega_{\rm b}$ and $n_s$ held fixed and using the a Gaussian prior of $h=0.738\pm0.024$ (R11), we find $q_0<0$ at 84 per cent confidence. The difference in constraints on the deceleration parameter can be understood as a result of the much larger values of the dark-energy density preferred by COSMOS-30 $\Omega_{\Lambda}=0.97^{+0.39}_{-0.60}$, which lead to smaller values of $q_0$. Whereas the dark-energy density found here from CFHTLenS is $\Omega_{\Lambda}=0.38\pm0.36$, resulting in larger values of $q_0$ for CFHTLenS. 

With CFHTLenS alone we are not able to put a strong constraint on the acceleration of the expansion of the Universe. With the addition of the other cosmological probes the entire posterior distribution of $q_0$ is less than zero. For a curved model with CFHTLenS combined with the other probes, we find $q_0=-0.594\pm0.014$ (see Table~\ref{tab:cosmo}). An accelerating Universe is unambiguously detected.

\section{Impact of non-linear effects and baryons on the tomographic cosmological constraints}
\label{sec:nonlinear}
We have presented cosmological parameter constraints from an analysis of the tomographic two-point shear correlation function $\xi^{\rm k,l}_\pm(\theta)$ (equation \ref{theogg}), incorporating the non-linear dark matter only power spectrum from \citet{Smith2003} as our theoretical model of $P_{\delta}(k,z)$ in equation (\ref{pofkappa}). Note that $k=\frac{\ell}{f_K(\chi)}$. This halo-model prescription for the non-linear correction has been calibrated on numerical simulations and shown to have an accuracy of $5-10$ per cent over a wide range of scales \citep{2011MNRAS.418..536E}. The N-body simulations used to estimate the covariance matrices  used in this analysis suggest that the accuracy is even better than this for a WMAP5 cosmology \citep{Clone2012} over the redshift range covered in this analysis. While these comparisons give us confidence in our results, and suggest that any error from the non-linear correction will be small in comparison to our statistical error, it is prudent to assess how errors in the non-linear correction will impact our results.

\begin{figure}
\centering
\includegraphics[scale=0.65,angle=0]{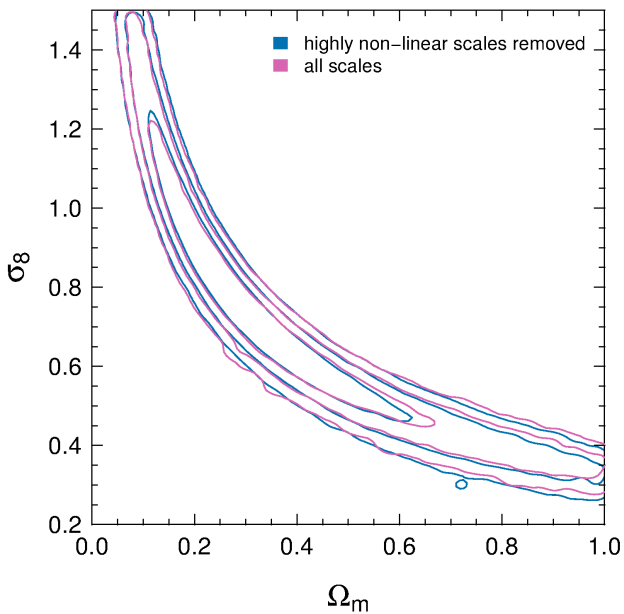}
\caption{\label{fig:comp}Marginalised parameter constraints (68.3, 95.5, and 99.7 per cent confidence levels) in the $\Omega_{\rm m}-\sigma_{\rm 8}$ plane for a flat $\Lambda$CDM cosmology. The pink contours show the result when all eight scales are included, this is the same as the result for CFHTLenS shown in Figure~\ref{fig:om_sig}. The blue contours show the result of removing highly non-linear scales, which are possibly biased due to the non-linear correction to the matter power spectrum or the effect of baryons. We remove the 5 smallest scales of $\xi_-$, corresponding to $\theta<10$ arcmin. The contours are only slightly different, indicating that we are not sensitive to these effects given the level of precision of our results.}
\end{figure}

A fully 3D weak lensing analysis of the CFHTLenS data is presented in \citet{Kitching2012}. This power spectrum analysis allows for exact redshift dependent cuts in the wave-vector $k$, which can be motivated by either the comparison of the power spectrum measured from N-body simulations to the non-linear prescription, or the selection of linear scales where the non-linear correction is negligible. For real-space statistics, as used in this paper and in \citet{Kilbinger2012}, it is not possible to make an unambiguous separation of scales. The two-point correlation function $\xi^{\rm k,l}_\pm(\theta)$ is related to the underlying matter power spectrum $P_{\delta}(k,z)$ through integrals over $k$ and $z$, modulated by the lensing efficiency $g(z)$ (equation~\ref{lensgeom}) and Bessel functions $J_{0/4}(\ell\theta)$ for $\xi_{+/-}(\theta)$ (see equations~\ref{theogg} and \ref{pofkappa}). The measured tomographic two-point shear correlation function $\xi^{\rm k,l}_\pm$ at a fixed scale $\theta$ is therefore probing a range of $k$ in the underlying matter power spectrum. In addition, owing to the different Bessel functions, $\xi_+$ is preferentially probing much smaller $k$, and hence larger physical scales, than $\xi_-$.

\citet{Kilbinger2012} present an analysis of the 2D shear correlation function out to large angular scales $\theta < 350$ arcmin. The consistent constraints obtained from the large quasi-linear regime $\theta > 53$ arcmin in comparison to the full angular range, analysed using the \citet{Smith2003} non-linear power spectrum, give us confidence that the accuracy of this correction is sufficient, falling within our statistical errors.

Comparing the theoretical expectation of $\xi^{\rm k,l}_\pm(\theta)$ (equation~\ref{theogg}) for a WMAP7 cosmology, calculated using a non-linear and a linear power spectrum, we determine the angular scale below which the non-linear and linear models differ in amplitude by greater than $10$ per cent. For $\xi_+$, this quasi-linear limit ranges from $10-14$ arcmin for the three different tomographic combinations (the lowest redshift bin requiring the largest $\theta$ cut). For $\xi_-$, the quasi-linear limit ranges from $100-140$ arcmin. In this analysis, we limit our angular range to scales with $\theta \lesssim 40$ arcmin where we can accurately assess a covariance matrix from the lensing simulations \citep{Clone2012}. We are therefore unable to follow \citet{Kilbinger2012} by limiting our real-space analysis to this quasi-linear regime as we do not probe sufficiently large angular scales. We can however make an assessment of how an error on the non-linear correction would impact our results. We first compare the WMAP7 theoretical expectation of $\xi^{\rm k,l}_\pm(\theta)$ calculated using a non-linear correction boosted by $7$ per cent, with a model calculated with the non-linear correction decreased by $7$ per cent. Note that we chose the value of $7$ per cent from the average error over the range of $k$ tested in \citet{2011MNRAS.418..536E}. We find that these two limits on the non-linear correction cause at least a $10$ per cent change in the amplitude of $\xi^{\rm k,l}_\pm(\theta)$ for scales $\theta \lesssim 1$ arcmin for $\xi_+$ and $\theta \lesssim 10$ arcmin for $\xi_-$. Applying these cuts in angular scale corresponds to removing the first 5 angular scales for $\xi_-$ for each tomographic bin shown in Figure~\ref{fig:signal}. All $\xi_+$ scales remain in the analysis. With these scales removed any remaining uncertainty due to the non-linear modelling is well within our statistical error.

Figure~\ref{fig:comp} compares cosmological parameter constraints in the $\Omega_{\rm m} - \sigma_8$ plane for this limited number of scales in comparison to the full data set analysed in Section~\ref{sec:tomography}. The removal of small scales results in a slight change to the degeneracy of the parameters. This test gives us confidence that the non-linear correction used is sufficiently accurate given the statistical error of the survey. This is unlikely to be true for future surveys, where the increased statistical accuracy will require better knowledge of the non-linear correction to the power spectrum \citep{2011MNRAS.418..536E}.

Finally we turn to the impact of baryons on our results. In our analysis we assume the underlying matter power spectrum is sufficiently well represented by the non-linear dark matter only power spectrum, neglecting the role of baryons. The impact of baryons on the power spectrum is sensitive to the baryonic feedback model used. Therefore, the magnitude of the impact of baryons remains uncertain. \citet{2011MNRAS.417.2020S} present an analysis of cosmological hydrodynamic simulations to quantify the effect of baryon physics on the weak gravitational lensing shear signal using a range of different baryonic feedback models. Their work suggests that a conservative weak lensing analysis should be limited to those scales where $k \lesssim 1.5 {\rm h\, Mpc^{-1}}$. We implement such a conservative scheme in the 3D power spectrum analysis of \citet{Kitching2012}. As discussed above, our real-space analysis mixes $k$ and $z$ scales, leaving us unable to perform a similarly clear test here. 

\citet{2011MNRAS.417.2020S} present a comparison of $\xi^{\rm k,l}_\pm(\theta)$ measured for both the cosmological hydrodynamic simulations and a dark matter only simulation for different redshifts, which we use to judge the level of error we should expect baryons to introduce. Assuming the realistic AGN feedback model, and considering the scales used in the conservative analysis of Figure~\ref{fig:comp} ($\theta \ge 1.34$ arcmin and $\theta \ge 15.4$ arcmin for $\xi_+$ and $\xi_-$ respectively), we expect baryons to cause a decrease to the modelled signal of less than ten per cent. The fact that we see very little difference between the conservative and full analysis presented in Figure~\ref{fig:comp} demonstrates that the underlying matter power spectrum is indeed sufficiently well represented by the non-linear dark matter only power spectrum for our statistical accuracy. It also indicates that the impact of baryons on the non-linear dark matter only power spectrum is unlikely to be larger than that predicted by \citet{2011MNRAS.417.2020S}. However, baryonic effects will have to be carefully considered for the next generation of weak lensing surveys that will have significantly smaller statistical errors. \citet{2012arXiv1210.7303S} show the importance of baryonic effects on three-point shear statistics and propose a modification to the modelling of the non-linear matter power spectrum to account for these effects.

\section{Conclusion}
\label{sec:conclusion}
The most important result of this study is that the sum of the photometric redshift probability distribution functions (PDF) within a redshift bin provides an accurate measure of the true redshift distribution of those galaxies; accounting for the scatter due to catastrophic as well as statistical errors. To demonstrate the accuracy of the PDFs we have compared the summed PDFs with the redshift distribution predicted by spectroscopic redshifts, resampled COSMOS-30 redshifts, and predictions from a redshift contamination analysis using the angular correlation function. We find excellent agreement for the redshift range $z_{\rm p}<1.3$. This result indicates that the priors and spectral templates used in \citet{Hendrick2012} to derive the photometric redshifts provide an accurate and complete description of the galaxies at $z_{\rm p}<1.3$. This also motivates our use of the summed PDF as a measure of the redshift distributions in our tomographic weak lensing analysis. Furthermore, the proven accuracy of the summed PDFs provides a reliable method for estimating the source redshift distribution in future weak lensing studies.

We have performed a cosmological analysis of the CFHTLenS data on angular scales $1<\theta<40$ arcmin, using two broad redshift bins, $0.5 < z_{\rm p} \leq 0.85$ and $0.85 < z_{\rm p} \leq 1.3$, that are not significantly affected by the intrinsic alignment of galaxy shapes. We model two cosmologies; flat and curved $\Lambda$CDM. Due to complementary degeneracies our results add valuable constraining power when combined with those from the cosmic microwave background \citep[][WMAP7]{WMAP7}, baryon acoustic oscillations \citep[][BOSS]{BOSS}, and a prior on the Hubble parameter \citep[][R11]{R11}. The addition of our weak lensing results to these other cosmological probes increases the precision of individual marginalised parameter constraints by an average factor of $1.5-2$.

For a flat $\Lambda$CDM model the joint parameter constraints for CFHTLenS, WMAP7, BOSS, and R11 are $\Omega_{\rm m}=0.2762\pm0.0074$ and $\sigma_{\rm 8}=0.802\pm0.013$. For a curved $\Lambda$CDM model, combining the same data sets, we find $\Omega_{\rm m}=0.2736\pm0.0085$, $\Omega_{\Lambda}=0.7312\pm0.0094$, $\Omega_{\rm K}=-0.0042\pm0.0040$, and $\sigma_{\rm 8}=0.795\pm0.013$. Full details of our parameter estimates for both cosmologies are presented in Table~\ref{tab:cosmo}. Our results are consistent with those presented in other studies of the CFHTLenS data: a 2D lensing analysis probing much larger scales where linear theory provides a more accurate model to the matter power spectrum \citep{Kilbinger2012}; and a fine-binned tomographic analysis with six redshift bins accounting for intrinsic alignments \citep{IIGIGG}. 

We compare the tomographic constraints with those from a 2D lensing analysis spanning the same range of redshift $0.5 < z_{\rm p} \leq 1.3$. We find the two analyses to be completely consistent with all parameter estimates agreeing within their 68.3 per cent confidence levels. We note that the uncertainties on individual parameters from tomography are on average 116 per cent larger than the uncertainties from 2D lensing. This statistic is 98.2 per cent if we replace our data vectors with a fiducial model, indicating that our covariance matrices do show a slight improvement for tomography. We argue that our non-Guassian covariance and broad overlapping redshift bins degrade the modest improvement (${\sim}88$ per cent) expected from idealised Fisher matrix calculations \citep{Simon2004}. We identify small scales as being responsible for 12 per cent of the increase from 98.2 per cent. These scales could be biased due to uncertainties in the modelling of the non-linearities in the matter power spectrum and baryonic effects. Although small scales have inflated our uncertainties from tomography we show in Section~\ref{sec:nonlinear} that they do not significantly affect our results.

Previous analyses of CFHTLS data were hindered by a strong redshift dependent bias in the weak lensing shear, necessitating additional nuisance parameters when analysing the tomographic shear \citep{MK09}. We demonstrate that the redshift scaling of the CFHTLenS cosmic shear signal agrees with expectations from the modelled $\Lambda$CDM cosmology. The strongest test of this is presented in Figure~\ref{fig:zscale}, which shows the agreement of cosmological constraints measured for each combination of redshift bins in the $\Omega_{\rm m}-\sigma_8$ plane for a flat $\Lambda$CDM cosmology. This demonstrates the effectiveness of the cosmology-independent tests of residual systematics presented in \citet{Heymans2012}, including the rejection of 25 per cent of the MEGACAM pointings which failed to pass these tests. Note also that the shear calibration performed on numerical simulations \citep{Miller2012} was completed before any cosmological analysis was performed on the data. The two-bin analysis presented here is sensitive to redshift dependent cosmology without introducing additional parameters to model intrinsic alignments, as such it is an excellent final test of the CFHTLenS data product.

\section*{acknowledgments}
This work is based on observations obtained with MegaPrime/MegaCam, a joint project of CFHT and CEA/DAPNIA, at the Canada-France-Hawaii Telescope (CFHT) which is operated by the National Research Council (NRC) of Canada, the Institut National des Sciences de l'Univers of the Centre National de la Recherche Scientifique (CNRS) of France, and the University of Hawaii. This research used the facilities of the Canadian Astronomy Data Centre operated by the National Research Council of Canada with the support of the Canadian Space Agency.  We thank the CFHT staff for successfully conducting the CFHTLS observations and in particular Jean-Charles Cuillandre and Eugene Magnier for the continuous improvement of the instrument calibration and the Elixir detrended data that we used. We also thank TERAPIX for the quality assessment and validation of individual exposures during the CFHTLS data acquisition period, and Emmanuel Bertin for developing some of the software used in this study. CFHTLenS data processing was made possible thanks to significant computing support from the NSERC Research Tools and Instruments grant program, and to HPC specialist Ovidiu Toader. The early stages of the CFHTLenS project were made possible thanks to the support of the European Commissions Marie Curie Research Training Network DUEL (MRTN-CT-2006-036133) which directly supported LF, HHi, BR, and MV between 2007 and 2011 in addition to providing travel support and expenses for team meetings.

The N-body simulations used in this analysis were performed on the TCS supercomputer at the SciNet HPC Consortium. SciNet is funded by: the Canada Foundation for Innovation under the auspices of Compute Canada; the Government of Ontario; Ontario Research Fund - Research Excellence; and the University of Toronto. 

LVW acknowledges support from the Natural Sciences and Engineering Research Council of Canada. CH and FS acknowledge support from the European Research Council under the EC FP7 grant number 240185. TE is supported by the Deutsche Forschungsgemeinschaft through project ER 327/3-1 and the Transregional Collaborative Research Centre TR 33 - "The Dark Universe". HHo and ES acknowledge support from  Marie Curie IRG grant 230924, the Netherlands Organisation for Scientific Research (NWO) grant number 639.042.814 and from the European Research Council under the EC FP7 grant number 279396. HHi is supported by the Marie Curie IOF 252760, a CITA National Fellowship, and the DFG grant Hi 1495/2-1. TDK acknowledges support from a Royal Society University Research Fellowship (NSERC) and the Canadian Institute for Advanced Research (CIfAR, Cosmology and Gravity program). YM acknowledges support from CNRS/INSU (Institut National des Sciences de l'Univers) and the Programme National Galaxies et Cosmologie (PNCG). BR acknowledges support from the European Research Council in the form of a Starting Grant with number 24067. TS acknowledges support from NSF through grant AST-0444059-001, SAO through grant GO0-11147A, and NWO. LF acknowledges support from NSFC grants 11103012 and 10878003, Innovation Program 12ZZ134 and Chen Guang project 10CG46 of SMEC, and STCSM grant 11290706600 and Pujiang Program 12PJ1406700. MJH acknowledges support from the Natural Sciences and Engineering Research Council of Canada (NSERC). MV acknowledges support from the Netherlands Organization for Scientific Research (NWO) and from the Beecroft Institute for Particle Astrophysics and Cosmology.

{\small Author Contributions: All authors contributed to the development and writing of this paper.  The authorship list reflects the lead authors of this paper (JB, LVW, CH, and MK) followed by two alphabetical groups.  The first alphabetical group includes key contributors to the science analysis and interpretation in this paper, the founding core team and those whose long-term significant effort produced the final CFHTLenS data product.  The second group covers members of the CFHTLenS team who made a significant contribution to either the project, this paper, or both.  The CFHTLenS collaboration was co-led by CH and LVW and the CFHTLenS Cosmology Working Group was led by TK.} 

\bibliographystyle{mn2e}
\bibliography{mn-jour}
\end{document}